\documentclass[print, amsmath,amssymb, aps,prc,twocolumn,nofootinbib]{revtex4-1}

\usepackage[usenames,dvipsnames]{xcolor}
\usepackage{graphicx}
\usepackage{dcolumn}
\usepackage{bm}
\usepackage{hyperref}

\newcommand{\nn}{\nonumber}
\renewcommand{\vec}[1]{{\bf #1}}

\newcommand{\etal}{\emph{et al.}}

\newcommand{\nuc}[2]{$^{#1}$\textrm{#2}}

\renewcommand{\Re}{\text{Re}}

\begin{document}


\title{Numerical accuracy of mean-field calculations in coordinate space}

\author{W. Ryssens}
\author{P.-H. Heenen}%
\email{phheenen@ulb.ac.be}
\affiliation{%
 PNTPM, CP229, Universit\'e Libre de Bruxelles, B-1050 Bruxelles, Belgium
}%
\author{M. Bender}
\affiliation{
  Universit\'e de Bordeaux, Centre d'Etudes Nucl\'eaires de Bordeaux Gradignan, UMR5797, F-33175 Gradignan, France
}%
\affiliation{
 CNRS/IN2P3, Centre d'Etudes Nucl\'eaires de Bordeaux Gradignan, UMR5797, F-33175 Gradignan, France
}%
\date{31 August 2015, compiled \today}

\begin{abstract}
\begin{description}
\item[Background] Mean-field methods based on an energy density functional (EDF) are powerful tools used to describe many properties of nuclei in the entirety of the nuclear chart. The accuracy required on energies for nuclear physics and astrophysics applications is of the order of 500~keV and much effort is undertaken to build EDFs that meet this requirement.

\item[Purpose] The mean-field calculations have to be accurate enough in order to preserve the accuracy of the EDF. We study this numerical accuracy in detail for a specific numerical choice of representation for the mean-field equations that can accommodate any kind of symmetry breaking.

\item[Method] The method that we use is a particular implementation of 3-dimensional mesh calculations. Its numerical accuracy is governed by three main factors: the size of the box in which the nucleus is confined, the way numerical derivatives are calculated and the distance between the points on the mesh.

\item[Results] We have examined the dependence of the results on these three factors for spherical doubly-magic nuclei, neutron-rich \nuc{34}{Ne}, the fission barrier of \nuc{240}{Pu} and isotopic chains around $Z = 50$.
\item[Conclusions] Mesh calculations offer the user extensive control over the numerical accuracy of the solution scheme. By making appropriate choices for the numerical scheme the achievable accuracy is well below the model uncertainties of mean-field methods.
\end{description}
\end{abstract}

\maketitle

\section{Introduction}

The self-consistent mean-field approach, based on an energy density functional (EDF), is a tool of choice to study nuclei in any region of the nuclear chart~\cite{bender03a}.
It allows to calculate the properties of the ground state but also of alternative configurations, like shape isomers, or to follow the behaviur of a nucleus along rotational bands or along fission paths.
Often, one is not directly interested in the total binding energy of a specific nucleus but by its evolution along a series of isotopes or isotones, which can signal structural changes for given neutron or proton numbers.

Motivated by the needs of the nuclear physics and astrophysics communities, large efforts are underway to push the predictive power of nuclear mass models well below the 500\,keV level.
To reach this goal, the protocols used to adjust the EDF's parameters have been revisited. In particular, methods are being developed~\cite{Gao13,Dob14a,JPGerrors} to quantify the statistical uncertainty of these parameters. However, besides the errors on observables due to these uncertainties, there is also a numerical error due to the way the SCMF equations are solved. One needs to verify that the numerics does not introduce errors that are larger than the maximum error tolerated for mass models. More importantly, these errors should not vary too rapidly from one nucleus to the other, to avoid a spurious behavior of mass differences.

The numerical methods used to solve the mean-field equations can be classified according to the way the single-particle wave functions are represented: by coordinate space techniques or by a basis expansion.
Coordinate space techniques represent the single-particle wave functions in a discretized, finite volume. Several discretization techniques exist, utilizing finite difference formulas~\cite{BFH85a}, Fourier transformations~\cite{Mar14a}, B-splines~\cite{Umar1991}, wavelets~\cite{PSF08a,Pei2014} and the Lagrange Mesh method~\cite{Bay86a,Imagawa03,Hashimoto13}.

The second family of numerical representations involves expanding the single-particle wave functions on some chosen (finite) set of basis states. Usually these basis states are harmonic oscillator (HO) eigenstates, although the details often vary.

While the origin of numerical errors is quite different for both families of representations, the type of EDF does not seem to influence the accuracy of the methods much. The three main families (relativistic EDFs, zero-range Skyrme EDFs or finite-range EDFs) require similar numbers of basis states to achieve a similar precision (see e.g.~\cite{Lu2014,HFBTHO1,Rodriguez2014,Schunck14,RER02a}). In what follows, we will limit ourselves to the study of zero-range Skyrme EDFs.

It is the aim of this paper to study the numerical accuracy of a specific implementation of coordinate space techniques: representation on a three-dimensional cartesian mesh of equidistant points. 
We will focus on two specific techniques: finite-difference (FD) formulas and the Lagrange Mesh (LM) method, which are the ones implemented in our codes. As far as we can infer from the tests published in the literature, the accuracy obtained with the other techniques mentioned above is similar to the one obtained within our LM scheme.
Most of the information relative to the tools that we have developed has been presented for the particular implementation made in the code \texttt{Ev8}~\cite{BFH05a,Ev8article}. More involved implementations have also been used, which differ from \texttt{Ev8} only in that they impose less symmetries on the nucleus. The presence of these symmetries in general allows for the reduction of the dimension of the problem, e.g.\ in \texttt{EV8} it allows for the reduction of the space by a factor of 1/8.

The article is organized as follows: first we define precisely the quantities that will be used to characterize the accuracy of a mean-field calculation. Next, we review the basic ingredients needed to define wave functions on a cartesian mesh and to calculate derivatives and integrals in this representation. We then discuss the main sources of numerical errors: the size of the box in which the nucleus is confined and the step size of the mesh. We discuss the numerical accuracy that can be achieved by comparing energies and radii of doubly magic nuclei with those obtained with a spherical code. Finally, we check the convergence of energies, radii and the multipole moments of deformed nuclei by comparing results obtained with decreasing mesh discretization lengths.

\section{Definition of useful quantities}

A mean-field configuration is characterized by its energy, its rms radius and by multipole moments. In this section we define these quantities whose dependence on the mesh parameters will be studied.

\subsection{Total Energy}
\label{sec:EnergyDef}
For a time-reversal invariant system as assumed here,
the total energy is composed of the kinetic energy, the
Skyrme energy describing the strong interaction in the particle-hole
channel, the pairing energy, the Coulomb energy and a center-of-mass
correction~\cite{Ev8article}
\begin{equation}
\label{eq:Etot}
E_{\text{tot}}
  =     E_{\text{kin}}
      + E_{\text{Skyrme}}
      + E_{\text{pair}}
      + E_{\text{Coul}}
      + E_{\text{cm}}
\, .
\end{equation}
For the parameterizations  used throughout this article, the Skyrme
EDF takes the form of the sum of various bilinear combinations of the
isoscalar ($t=0$) and isovector ($t=1$) local densities $\rho_t(\vec{r})$,
kinetic densities $\tau_t(\vec{r})$ and spin-current densities
$J_{t,\mu \nu}(\vec{r})$, $\mu$, $\nu = x$, $y$, $z$,
\begin{eqnarray}
\label{eq:skyrme:energy}
E_{\text{Skyrme}}
& = &   E_{\rho^2}
      + E_{\rho^{2 + \alpha}}
      + E_{\rho \tau}
      + E_{\rho \Delta \rho}
      + E_{\rho \nabla J}
      + E_{JJ}
      \nonumber \\
& = & \sum_{t=0,1} \int \! d^3r \,
      \Big(
               C^\rho_t [\rho_0] \, \rho_t^2
             + C^{\rho^{\alpha}}_t \rho_0^\alpha \, \rho_t^2
             + C^\tau_t          \, \rho_t \, \tau_t
      \nonumber \\
&   &
             + C^{\Delta \rho}_t \, \rho_t \, \Delta \rho_t
             + C^{\nabla \cdot J}_t  \, \rho_t \nabla \cdot \vec{J}_t
      \nonumber \\
&   &
             - C^{T}_t \sum_{\mu, \nu} J_{t, \mu \nu} \, J_{t, \mu \nu}
      \Big) \, ,
\end{eqnarray}
with coupling constants as defined in Ref.~\cite{Ev8article}.
The kinetic energy just depends on the kinetic density
\begin{equation}
\label{eq:E_kin}
E_{\text{kin}}
= \sum_{q = n,p} \int \! d^3r \; \frac{\hbar^2}{2m_q} \, \tau_q (\vec{r})\, .
\end{equation}
of protons and neutrons. While the Skyrme and kinetic energies are
local functionals of the densities, the direct Coulomb energy is a nonlocal
functional of the proton density $\rho_p(\vec{r})$
\begin{equation}
\label{eq:Ecoul}
E_{\rm Coul}^d
= \frac{e^2}{2}
  \iint \! d^3r \, d^3r^\prime \;
  \frac{\rho_p(\vec{r}) \, \rho_p(\vec{r^\prime})}
       {|\vec{r}-\vec{r^\prime|}}
\, .
\end{equation}
Compared to the other terms contributing to the total energy \eqref{eq:Etot},
the exact calculation of the Coulomb exchange energy is orders of magnitude
more costly as it is a functional of the complete nonlocal one-body density
matrix. As a consequence, the local Slater approximation that is of similar
numerical cost as the Skyrme energy \eqref{eq:skyrme:energy} is used instead
\begin{equation}
\label{eq:Slater}
E_{\rm Coul}^{e}
= - \frac{3 e^2}{4} \left( \frac{3}{\pi} \right)^{1/3}
  \int \! d^3r \; \rho^{4/3}_p (\vec{r}) \, .
\end{equation}
The pairing energy contribution to the energy is:
\begin{equation}
E_{\text{pair}}
 =  \sum_{k, m > 0} f_k \, u_k v_k \, f_m \, u_m v_m \,
      \bar{v}^{\text{pair}}_{k \bar{k} m \bar{m}}
\, ,
\end{equation}
where the $\bar{v}_{k \bar{k} m \bar{m}}$ are anti-symmetrized matrix
elements of the pairing interaction and the $f_i$ are cutoff factors, both of which are specified in Appendix~\ref{app:pairing}.

The expression for the c.m.\ correction, which is not relevant for our discussion, can be found in Ref.~\cite{Ev8article}.

\subsection{Dimensionless multipole moments}
As in \cite{Ev8article}, the dimensionless multipole moments $\beta_{\ell m}$ are related to the matrix elements of the multipole operators  $\hat{Q}_{\ell m}\equiv r^\ell \, Y_{\ell m}(\vec{r})$ by
\begin{equation}
\label{equ:betalm}
\beta_{\ell m}
=  \frac{4 \pi}{3 R_0^\ell A} \langle \hat{Q}_{\ell m} \rangle \, ,
\end{equation}
where $R_0 = 1.2 \, A^{1/3} \, \text{fm}$. When $m$ is omitted we imply it to be zero.

\subsection{Radii}

Another set of observables, related to the density profile of the nucleus, are the mean-square (ms) radii, root-mean-square (rms) radii and the isotopic shifts. The ms radius of the proton ($q= \text{p}$), neutron ($q= \text{n}$), and total density distribution is defined as
\begin{eqnarray}
r^2_{q}
&=& \frac{1}{N_q} \int \! d^3r \; \rho_q(\vec{r} ) \, r^2 \, ,\\
r^2_{t} 
&=& \frac{1}{A} \int \! d^3r \; \left[ \rho_n( \vec{r} ) + \rho_p( \vec{r} ) \right] \, r^2 \, .
\end{eqnarray}
The root-mean-square (rms) radii are then the square root of the corresponding
mean-square radius.

Similarly, we will present results for the isotope shift of charge radii that are calculated as the difference between the proton ms radius of an isotope with $N$ neutrons and a reference isotope with $N_0$ neutrons
\begin{equation}
\delta r^{2} (N,Z) 
= r^2_{p}(N,Z) - r^2_{p}(N_0,Z)
\end{equation}
without any corrections.

\section{Coordinate space representation}

Assuming a 3-dimensional cartesian mesh, a function $\Phi(\vec{r}) = \Phi(x,y,z)$
is represented by the tensor $\Phi_{pqs}$ of its values at
the collocation points $(x_p, y_q, z_s)$
\begin{equation}
\Phi(\vec{r})
= \{ \Phi(x_p, y_q, z_s) \}
= \{ \Phi_{pqs} \} \, .
\end{equation}
A mesh can be defined in several ways, depending on
the choice of the collocation points. For example,
the origin of the coordinate system and the boundaries of the box can be included as collocation points or not. Different choices
can also be made for the boundary conditions at the edge of the box.

To set up the self-consistent mean-field equations, one has to vary the EDF with respect to $\Phi_{pqs}$. This requires to define
prescriptions to calculate derivatives and integrals from the values of $\Phi_{pqs}$ on the mesh.
Several choices for derivatives have been explored over the years.

%
%

\subsection{Derivatives on a mesh}
\label{sec:deronamesh}

The most straightforward possibility to set-up a coordinate-space
representation of the self-consistent mean-field equations is provided
by the finite-difference method, a widely-used tool to solve
partial differential equations~\cite{Olver}.

In such a scheme, the derivatives are calculated with $n$-point
finite-difference formulas, and the integrals are obtained by summing
up the integrand at the mesh points multiplied by a suitable volume
element.

There are three factors that determine the accuracy that can be achieved
with the finite-difference method. One is the overall resolution scale
provided by the mesh spacing; decreasing the distance between mesh points
improves the accuracy. Second, the higher the order of the finite-difference formulas
used for a given mesh spacing the better the accuracy.  In both cases, however, better accuracy means also
increase of the numerical cost. Third, there are internal inconsistencies
introduced by the method itself. For example, taking twice the numerical
first-order derivatives of a given function is not equivalent to applying
the numerical second-order derivatives. Also, the numerical derivatives
are not the inverse of the numerical integration. It is only for very small
step sizes well below 0.1\,fm that these internal inconsistencies
become irrelevant.
While such small step sizes can be easily handled in spherical 1d codes~\cite{Karim}, the required storage is prohibitive in
axial 2d and cartesian 3d codes. In addition, such step sizes are much
smaller than what can be expected to be the physically relevant resolution
scale, see for example the arguments brought forward in Ref.~\cite{Bul13a}.

Several other
schemes have been developed in the past with a better consistency between derivation and integration.
For instance, derivatives have been defined through a Fourier transformation
to momentum space~\cite{Blu92a,RutzThesis,Mar14a}, which is equivalent to the assumption that the functions on the mesh can be developed into
a set of plane waves. In this method, the derivatives are quasi-exact
for a given resolution of the mesh, and first- and second-order
derivatives internally consistent. Similar ideas have been developed
in quantum chemistry under the label of \textit{discrete variable
representation} (DVR)~\cite{Sza99a,Sza12a,Lit02b}. A similar formalism
that provides an internal consistent scheme for derivatives
and integrals is the Lagrange-mesh method that we will sketch in the
following section.
%
%

\subsection{Lagrange-mesh representation}

The idea underlying the Lagrange-mesh method is that for each
Gauss quadrature one can construct a set of basis functions for which
orthogonality and completeness relations are exactly fulfilled when
evaluated with the given quadrature~\cite{Bay86a,Bay06a,Bay15a}.
This additional condition makes the LM method a special case of the
slightly less rigorous concept of DVR~\cite{Bay15a,Sza12a}.

Lagrange meshes have been constructed for a multitude of different
geometries and used for a wide range of applications, see~\cite{Bay15a}
and references therein.
We will use here the case of an equidistant
3d cartesian mesh. Its three directions are separable in the formalism,
such that the presentation of the principles of the method for one
dimension is sufficient.

The underlying basis of a one-dimensional Cartesian equidistant Lagrange
mesh is constructed as the set of functions $\varphi_{k}(x)$ whose
orthogonality relations are exact when evaluated with a simple $2N$-point
rectangular quadrature rule, sometimes called \textit{midpoint rule},~\cite{Bay86a}
\begin{eqnarray}
\lefteqn{
\int_a^b \! dx \; \varphi^*_{k} (x) \, \varphi_{k'} (x)
} \nn \\
& \to & dx
        \sum_{r}
        \varphi^*_{k} (x_r) \, \varphi_{k'} (x_r)
  =     \delta_{k k'} \, ,
\end{eqnarray}
where $dx$ is the distance between the collocation points located at
\begin{equation}
 x_r = r \, dx = \pm dx/2, \pm 3 \, dx/ 2, \ldots, \pm (N-1) \, dx/ 2 \, ,
\end{equation}
and where $a$ and $b$ are the boundaries of the numerical box
$[a,b] = [-N dx, +N dx]$.
A convenient representation of the $2N$ basis functions $\varphi_{k}(x)$
are plane waves of the form
\begin{equation}
\label{eq:mash:cart:basis:1}
\varphi_{k}(x)
= \frac{1}{\sqrt{L}}
  \exp \Big( \tfrac{2\pi i}{L} k \, x \Big) \, ,
\end{equation}
where $L = 2 N dx$ is the length of the numerical box and where
$k = \pm \tfrac{1}{2}$, $\pm \tfrac{3}{2}$, \ldots, $\pm(N-\tfrac{1}{2})$.
The real part of the $\varphi_{k} (x)$ is symmetric, has nodes on
the boundaries of the box and a maximum at the origin, whereas their
imaginary part is skew-symmetric, consequently has a node at the origin,
and maxima on the boundaries of the box. This also implies that
$\varphi^*_{k} (x) = \varphi_{-k} (x)$.

The $\varphi_{k}(x_r)$ form a complete set of functions to describe
any function on the mesh points
\begin{eqnarray}
dx \sum_{k}
\varphi^*_{k} (x_r) \, \varphi_{k} (x_s)
& = & \delta_{rs}
\, .
\end{eqnarray}
Note that the box size $L$ is not a multiple of the wavelength
of the basis functions. Instead, twice the box size is an odd multiple
of the wavelengths that take the values $2L = 2L/1$, $2L/3$, $2L/5$,
\ldots $2L/(2N - 1)$. Both the real and imaginary parts of all plane waves in 
Eq.~\eqref{eq:mash:cart:basis:1} are non-zero at all mesh points.

As recalled in Ref.~\cite{Bul13a}, in cartesian DVR and LM
coordinate-space methods
where the derivatives are defined through an expansion in plane waves,
the analysis of a calculation's infrared and ultraviolet cutoffs
introduced by the basis is straightforward. This has to be
contrasted with the much more involved analyses required when working
with an HO basis~\cite{Fur12a,Coo12a,Fur15a}. It has also been argued in
Ref.~\cite{Bul13a} that a DVR or LM representation of the nuclear
many-body problem covers the relevant part of the phase space with a much
smaller number of basis states than required by an HO basis. In
practice, however, HO bases typically used for self-consistent mean-field
calculations are much smaller than the typical number of mesh points used
in the same kind of calculation. For a box with 20 points in every direction
the number of linearly independent states is 64000, to be compared with a
harmonic oscillator expansion with 20 shells, which contains 14168 states.

While the basis functions $\varphi_k (x)$ of Eq.~\eqref{eq:mash:cart:basis:1}
are useful to discuss the mathematical properties of the LM method, the
actual coordinate representation then employs the set of $2N$ Lagrange
interpolation functions $f_i(x)$ obtained as~\cite{Bay86a,Bay06a,Bay15a}
\begin{eqnarray}
\label{lag_f}
f_r (x)
& \equiv & dx
      \sum_{k}
      \varphi^*_k (x_r) \, \varphi_k (x)
      \nn  \\
& = & \frac{1}{2N}
      \frac{\sin \big[\frac{\pi}{dx} (x-r \, dx)\big]}
           {\sin \big[\frac{\pi}{dx}\frac{x-r \, dx}{2N}\big]}
\, .
\end{eqnarray}
By construction, the Lagrange interpolation functions have the property to be equal to one at the mesh point $x_r = r \, dx$, and zero at all others, $f_r (x_s) = \delta_{rs}$\cite{Bay86a,Bay06a,Bay15a}. When developed into the Lagrange functions, any function $\phi(x)$ on the mesh
\begin{equation}
\label{eq:discret}
\phi(x)
= \sum_{r}
  \phi (x_r) \, f_r(x)
= \sum_{r}
  \phi_r \, f_r(x)
\end{equation}
is then simply represented by its values $\phi_r \equiv \phi (x_r)$
at the $2N$ mesh points.

The Lagrange functions are smooth and infinitely derivable. They
can be used to define matrices representing the first and second derivatives
of functions discretized through\footnote{Unfortunately, the corrections of 
these expressions as given in the corrigendum to Ref.~\cite{Ev8article} still contain a typographical error: the formula for the second derivative has a superfluous factor of two when $i \ne j$.}
 Eq.~\eqref{eq:discret} 
\begin{widetext}
\begin{eqnarray}
\label{eq:der1}
D^{(1)}_{ji}
& \equiv & \frac{d f (x)}{dx} \bigg|_{x=x_j}
      \nn \\
& = & \left\{ \begin{array}{ll}
          {\displaystyle
          (-1)^{i-j} \, \frac{\pi}{(2N)dx} \, \frac{1}{\sin(\pi (i-j)/(2N))} } &
          \text{for $i \neq j$,} \\
          0 &
          \text{for $i = j$,}
          \end{array} \right.
          \\
\label{eq:der2}
D^{(2)}_{ji}
& \equiv & \frac{d^2 f_i(x)}{dx^2}\bigg|_{x=x_j}
      \nn \\
& = & \left\{ \begin{array}{ll}
          {\displaystyle
           (-1)^{i-j+1} \, 2 \, \left(\frac{\pi}{(2N)dx}\right)^2
           \frac{\cos \left[\pi (i-j)/(2N)\right]}{\sin^2\left[\pi (i-j)/(2N)\right]}
           }
           & \text{for $i \neq j$}, \\
           {\displaystyle
           - \frac{\pi^2}{3dx^2} \left(1-\frac{1}{(2N)^2}\right) }
           & \text{for $i = j$.}
          \end{array} \right.
\end{eqnarray}
\end{widetext}
The first derivative of any function $\phi(x)$ on the mesh is
obtained by multiplying the $2N \times 2N$ matrix $D^{(1)}_{rs}$
by the vector $\phi_r$
\begin{eqnarray}
\phi^{\prime}_r
& = & \phi^{\prime} (x_r)
  =   \sum_{s} 
      D^{(1)}_{rs} \, \phi_s \, ,
\end{eqnarray}
and similar for the second derivatives. Note that the derivative matrices
have the property $D^{(2)} = D^{(1)} D^{(1)}$ by construction~\cite{Bay15a},
which is not the case for finite-difference formulas. As the derivatives of Eqs.~\eqref{eq:der1} and \eqref{eq:der2} correspond to full $2N \times 2N$ matrices, their application is more time consuming than finite-difference derivatives
that correspond to a sparse band matrix.

The full cartesian 3d representation of a function $\Phi(\vec{r})$ is then
provided by
\begin{equation}
\Phi(\vec{r})
= \sum_{pqs} \Phi_{pqs} \, f_p(x) \, f_q(y) \, f_s(z) \, ,
\end{equation}
where the number of
discretization points does not have to be the same in each direction. In that case, the derivative
matrices in Eqs.~\eqref{eq:der1} and \eqref{eq:der2} have to be set-up
separately for each direction.

As pointed out in Refs.~\cite{Bay02a,Sza12a}, a variational calculation
using a DVR or LM derivatives delivers very precise values for the total energy
in spite of the individual matrix elements being much less accurate.
In what follows, we will illustrate that this property implies very accurate total energies while separate terms
of the Skyrme EDF are less well represented. In addition, we will show that using a LM results in a variational calculation.

%
%
\section{Numerical considerations}

\subsection{Numerical parameters of parameterizations}

Unless explicitly stated, we have used the SLy4 parameterization. To explore the dependence of the numerical accuracy on the EDF, we have in addition tested a representative set of Skyrme parameterizations, as listed in Appendix~\ref{app:Forces}.

In the next two sections, we present calculations for doubly-magic spherical nuclei $^{40}$Ca, $^{132}$Sn and $^{208}$Pb, the neutron-rich nucleus $^{34}$Ne, Cd, Sn and Te isotope chains and the fission path of $^{240}$Pu. It is worth noting that we only included pairing for the isotopic chains and for $^{240}$Pu, see appendix~\ref{app:pairing} for details. In all other cases, pairing has been neglected.

In appendix~\ref{app:Constants} we comment on the precise physical constants used during our calculations.

\subsection{Measuring accuracy}

The accuracy of a coordinate space calculation is limited by the size of the box, the discretization length $dx$ and the way derivatives and integrals are calculated. In order to properly judge these effects we employ two ways of analyzing results. For spherical nuclei we can compare our 3d results with a one-dimensional spherical code that also represents the single-particle wave functions in coordinate space. Because of spherical symmetry, we can use extremely fine discretizations and the results can thus be considered exact to very high precision. For this purpose we use \texttt{Lenteur}~\cite{Karim} as a reference.

For deformed nuclei, we no longer have access to such a comparison. Here we have to resort to looking at 3d results as a function of both box size and mesh spacing: we compare results in small boxes with large mesh spacing to results in very large boxes with very fine mesh spacing.

\subsection{The use of derivatives and the variational principle}

\begin{figure}
 \includegraphics[width=.45\textwidth]{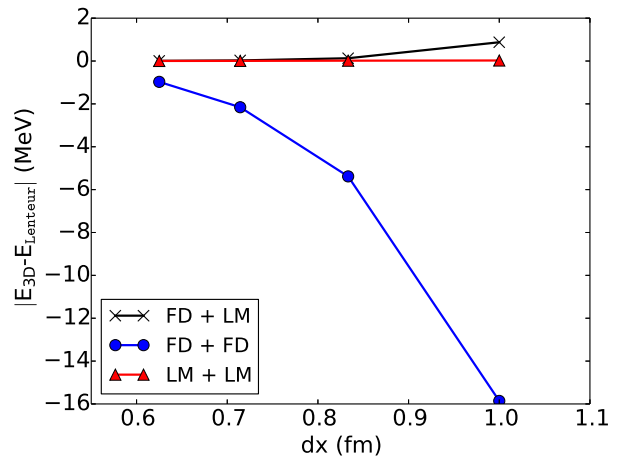}
 \caption{(Color online) Comparison between the errors on the total energy of \nuc{208}{Pb} obtained in calculations using different combinations of formulas for the derivatives (see text). The differences are taken with respect to the total energy obtained with \texttt{Lenteur} \cite{Karim}.}
 \label{fig:NewOldComparison}
\end{figure}

\begin{table*}
 \begin{tabular}{lddddd}
  \hline\hline\noalign{\smallskip}
  Energy (MeV)               & \multicolumn{1}{c}{$dx = 1.0$\,fm} & \multicolumn{1}{c}{$dx = 0.83$\,fm} & \multicolumn{1}{c}{$dx = 0.71$\,fm} & \multicolumn{1}{c}{$dx = 0.549$\,fm}  & \multicolumn{1}{c}{\texttt{Lenteur}}\\
  \noalign{\smallskip}\hline\noalign{\smallskip}
  Kinetic + c.m.~correction  & 3908.548   & 3880.647   & 3872.035 &  3867.506  &  3866.190 \\
 \noalign{\medskip} 
  Direct Coulomb             & 831.801    & 829.241    &  828.433 &   828.004    &  827.876  \\
  Coulomb Exchange           & -31.415    & -31.319    &  -31.289 &   -31.273    &  -31.269  \\
 \noalign{\medskip} 
  $E_{\rho^2}$               & -22749.578 & -22510.048 & -22435.391 & -22395.941 & -22384.379 \\
  $E_{\rho \tau}  $          &   1368.924 &   1343.506 &   1335.611 &   1331.436 &   1330.206 \\
  $E_{\rho^{2 + \alpha}}$    &  14812.851 &  14631.848 &  14575.413 &  14545.584 &  14536.832 \\
  $E_{\rho \Delta \rho}$     &    323.771 &    318.147 &    316.431 &    315.539 &    315.287 \\
  $E_{\rho \nabla J}$        &    -99.730 &    -97.595 &    -96.917 &    -96.554 &    -96.445 \\
 \noalign{\medskip} 
  Total Skyrme Energy        & -6343.762  & -6314.142  &  -6304.854 & -6299.937  & -6298.501 \\
 \noalign{\medskip} 
  Total Energy               & -1634.828  & -1635.574  &  -1635.675 & -1635.700 & -1635.703 \\
  \noalign{\smallskip}\hline\hline
 \end{tabular}
\caption{Decomposition of the total energy between the terms of the Skyrme parametrization SLy4 for \nuc{208}{Pb}, using the FD+LM option. All energies are in MeV. See Sect.~\ref{sec:EnergyDef} for the definition of the various terms. Values obtained with the spherical 1d code \texttt{Lenteur} are given for comparison.}
\label{tab:EDFComp}
\end{table*}

\begin{table*}
 \begin{tabular}{lddddd}
  \hline\hline\noalign{\smallskip}
  Energy (MeV)               & \multicolumn{1}{c}{$dx = 1.0$\,fm} & \multicolumn{1}{c}{$dx = 0.83$\,fm} & \multicolumn{1}{c}{$dx = 0.71$\,fm} & \multicolumn{1}{c}{$dx = 0.549$\,fm} & \multicolumn{1}{c}{\texttt{Lenteur}}\\
  \noalign{\smallskip}\hline\noalign{\smallskip}
  Kinetic + c.m.~correction  & 3866.323   & 3866.165   &  3866.182  & 3866.182 &  3866.190  \\
 \noalign{\medskip} 
  Direct Coulomb             &  827.922   & 827.889    &  827.882   &  827.878 &  827.876   \\
  Coulomb Exchange           & -31.269    & -31.268    &  -31.268   &  -31.268 &  -31.269    \\
 \noalign{\medskip} 
  $E_{\rho^2}$               & -22384.936 & -22384.188 & -22384.322 & -22384.320 & -22384.379  \\
  $E_{\rho \tau}  $          &   1330.300 &   1330.193 &   1330.201 &   1330.200 &   1330.206 \\
  $E_{\rho^{2 + \alpha}}$    &  14537.174 &  14536.691 &  14536.789 &  14536.787 &  14536.832 \\
  $E_{\rho \Delta \rho}$     &    315.238 &    315.275 &    315.284 &    315.284 &    315.287 \\
  $E_{\rho \nabla J}$        &    -96.430 &    -96.444 &    -96.445 &    -96.445 &    -96.445 \\
 \noalign{\medskip} 
  Total Skyrme Energy        & -6298.657  & -6298.473  &  -6298.493 &  -6298.494 &  -6298.501 \\
 \noalign{\medskip} 
  Total Energy               & -1635.678  & -1635.687  &  -1635.696 & -1635.700  &  -1635.703 \\
  \noalign{\smallskip}\hline\hline
 \end{tabular}
\caption{\label{tab:EDFCompL}
Same decomposition as in Table~\ref{tab:EDFComp}, but for the LM+LM option.}
\end{table*}

The numerical cost of using LM derivatives is much higher than the FD alternative.
 To control the computational time, three options have been considered: they differ by the way derivatives are calculated during the mean-field iterations and after convergence. The first option (FD+FD) has been used in the first applications of the codes \cite{BFH85a} where derivatives were exclusively calculated by FD. The second one (FD+LM) is the most used one for more than 20 years: FD derivatives are used during the iterations
 but the energies are recalculated after convergence by LM formulas. Finally, in the last option (LM+LM), the LM formulas are used during the iterations and after convergence.

In practice, we use a seven-point difference formula for the first order and a nine-point formula for the second order derivatives when employing FD formulas. It has been shown earlier in Ref.~\cite{Heenen93} that this provides an efficient compromise in terms of overall speed and precision.

Figure~\ref{fig:NewOldComparison} illustrates the accuracy on the total energy obtained using these three options. The LM + LM choice is by far the most accurate. As it can be seen in Table~\ref{tab:EDFCompL},
the result obtained with a mesh size of 1.0~fm differs by only 25~keV from the \texttt{Lenteur} result. The FD+LM option is less accurate, but already sufficient for most applications, with an error of around 100~keV for $dx=0.8$~fm. It is better by nearly an order of magnitude than the FD+FD choice. Results presented in the following have been obtained with the FD+LM option, except otherwise stated.

Both the FD+LM and LM+LM calculations underestimate the binding energy, as it should be for a variational calculation. This is due to the fact that the single-particle wave functions are expanded on a complete and closed  basis for given box size and mesh discretization length, see Eq.~\eqref{eq:mash:cart:basis:1}. Increasing the box size and/or decreasing the mesh discretization length  enlarge the accessible subspace of the Hilbert space \cite{Bay15a} and lead to a monotonous convergence of the energy. By contrast, such a basis cannot be defined for the FD+FD option for which the calculation systematically overestimates the binding energy of \nuc{208}{Pb}.

The same applies to mesh calculations with Fourier derivatives, as can be deduced from the convergence analyses in Refs.~\cite{Blu92a,Mar14a}. While for a given $dx$ the overall accuracy of the binding energy found there is very similar to the one we find for LM+LM calculations, the energy does not converge monotonically when decreasing $dx$.

While the use of LM derivatives after having used FD ones during the iterations (FD+LM) is sufficient to obtain an upper bound of the total energy since any wave function discretized on a mesh can be expanded on the LM basis,
the errors on the various individual terms of the Skyrme EDF can be very large, as can be seen in Table~\ref{tab:EDFComp}. While the total energy varies by slightly less than one MeV when $dx$ is decreased from $1.0$~fm to $0.549$~fm, the variation in the kinetic energy is of the order of 40~MeV, counterbalanced by a similar change in the Skyrme energy. The situation for the LM+LM scheme is shown in Table~\ref{tab:EDFCompL}. It indicates a similar effect, but on a much smaller scale: the total energy varies by 20~keV while the kinetic energy varies by roughly 150~keV.

When performing symmetry restoration and configuration mixing by the GCM, a high level of accuracy is required to avoid buildup numerical noise while solving the Hill-Wheeler-Griffin equation. This calls for the use of LM derivatives in these calculations, as done since our first applications~\cite{HBD93a}.

\subsection{Determining box sizes and mesh spacings}
\label{sec:boxsizeandmeshspacing}

\begin{figure}
 \includegraphics[width=.45\textwidth]{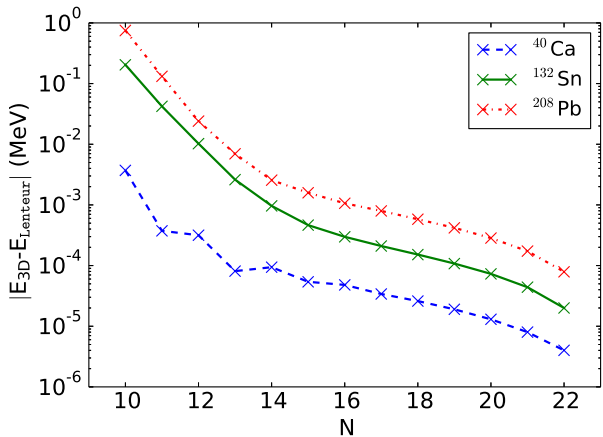}
 \caption{(Color online) Energy difference between a reference calculation performed with 23 points and calculations performed with $N$ points for the three nuclei \nuc{40}{Ca}, \nuc{132}{Sn} and \nuc{208}{Pb}. In all cases, the step size is equal to 1.0~fm.}
 \label{fig:BoxSizeSphericals}
\end{figure}

The first requirement of a coordinate space calculation is that the box in which the nucleus is confined is large enough to avoid any spurious effect due to the truncation of the wave functions. The
influence of the box size on the total energy for three spherical nuclei is represented in Fig.~\ref{fig:BoxSizeSphericals}. The same mesh size $dx=1.0$ fm is used in all calculations while the number of discretization points is varied, changing thus the volume of the box. The calculation in the largest box, using 23 points, is taken as a reference. The errors decrease quickly when the box size is enlarged. If one requires that the error is smaller than a keV, we see that taking boxes with half-sides of $11$ fm for \nuc{40}{Ca}, $15$ fm for \nuc{132}{Sn} and $20$ fm for \nuc{208}{Pb} is sufficient. Since the numerical effort required for $^{40}$Ca is very low we opted to use a slightly larger half-side of $13$ fm in order to further increase our accuracy to about 0.1 keV.

\begin{table}
 \begin{tabular}{cdddd}
  \hline\hline\noalign{\smallskip}
  nuclei & \multicolumn{1}{c}{$L_x = L_y$ }& \multicolumn{1}{c}{$L_z$} &  \multicolumn{1}{c}{$C_x = C_y$} & \multicolumn{1}{c}{$C_z$} \\
  \noalign{\smallskip}\hline\noalign{\smallskip}
  \nuc{40}{Ca}      & 26   & 26   & 26   & 26 \\
  \nuc{132}{Sn}     & 30.8 & 30.8 & 46.8 & 46.8 \\
  \nuc{208}{Pb}     & 40   & 40   & 60   & 60 \\
  {$Z\approx 50$}   & 40   & 40   & 60   & 60 \\
  \nuc{240}{Pu}     & 40   & 60   & 80   & 120 \\
  \noalign{\smallskip}\hline\hline
   \end{tabular}
 \caption{Edge lengths in fm of the boxes used to solve the self-consistent
mean-field equations ($L_\mu$) and to determine the Coulomb potential 
($C_\mu$) for the nuclei studied in this paper.
Depending on the symmetries imposed on the nucleus, only half of of the 
length is treated numerically in most cases.
}
\label{tab:boxsize}
\end{table}

Similar analyses have been performed for all nuclei considered in this paper. Since several nuclei in the isotopic chains around $Z=50$ are deformed, we have performed all calculations with the same box size
as \nuc{208}{Pb}. This choice allows us to calculate all isotopes with the same numerical conditions.
The box dimensions are summarized in Table \ref{tab:boxsize}. The columns $C_x, C_y$ and $C_z$ indicate the size of the box in which the Coulomb problem is solved. For every system, the box size was varied for fixed $dx$ until the energy did not change by more than $0.1$~keV, with the exception of the \nuc{240}{Pu} for which this limit was of 1~keV.

A non-ambiguous comparison between calculations performed with different mesh discretizations $dx$ can only be achieved when the volume of the box is conserved.  This is realized by determining the value of $dx$ in such a way that the box has the same size for each number of mesh points.

\begin{figure}[t!]
 \includegraphics[width=.45\textwidth]{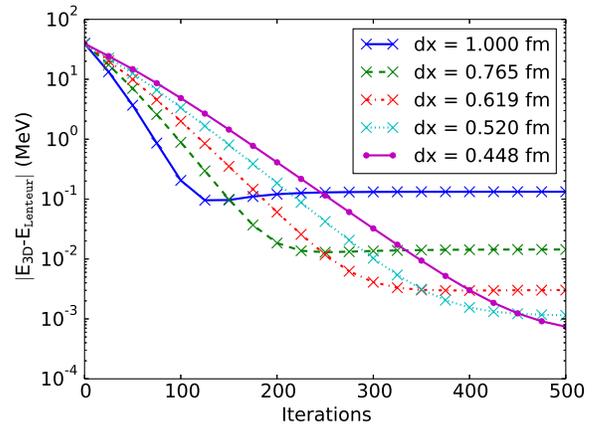}
 \caption{(Color online) Error on the total energy using the FD+LM option as a function of the number of mean-field iterations for \nuc{40}{Ca} for different values of $dx$. Calculations were initialized with Nilsson-model single-particle wave functions. The box length is $26$ fm.}
 \label{fig:Convergence}
\end{figure}

\begin{figure}[t!]
 \includegraphics[width=.45\textwidth]{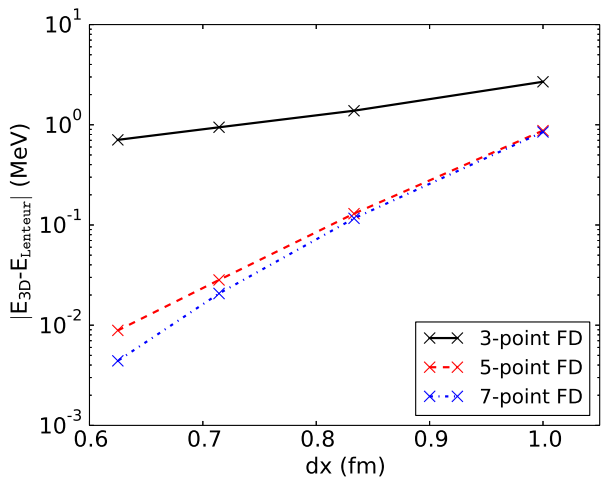}
 \caption{(Color online) Differences between the total energy of \nuc{208}{Pb} calculated on a 3d mesh with different approximations in the calculation of the Coulomb energy and that obtained with \texttt{Lenteur}. The three lines correspond to using 3-, 5- and 7-point FD formulas for the calculation of the Laplacian in the Poisson equation.}
  \label{fig:coulcomparison}
\end{figure}

\begin{figure*}[t!]
 \includegraphics[width=.9\textwidth]{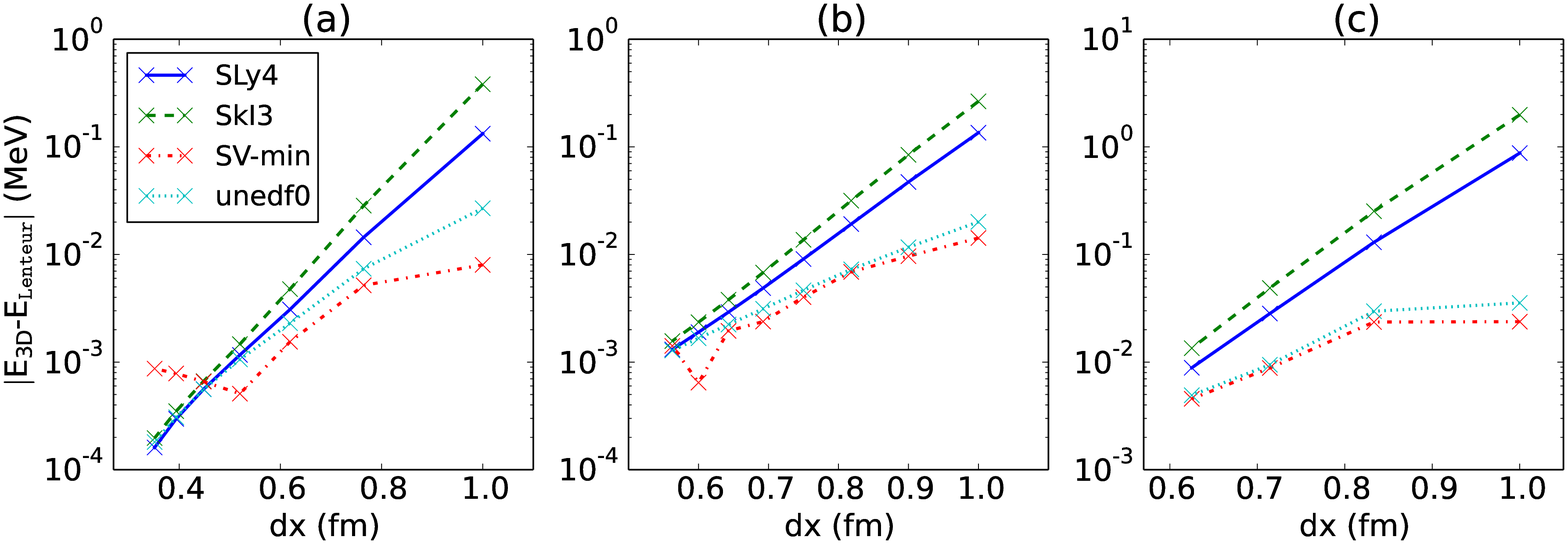}
 \caption{(Color online) Differences between the total energy calculated with our 3-dimensional code and with the spherical code \texttt{Lenteur}  for \nuc{40}{Ca} (a), \nuc{132}{Sn} (b) and \nuc{208}{Pb} (c), as a function of mesh distance $dx$. Results are plotted for a representative set of Skyrme parameterizations, without pairing. Results for SLy5, T22 and T65 are not shown, but are indistinguishable from the SLy4 results on the scale of this graph.}
\label{fig:DxSphericals}
\end{figure*}

\subsection{Convergence of the iterative procedure}

Decreasing the mesh size improves the accuracy. However, this has a price in computing time. First, to keep the same box size requires to increase the number of discretization points. A second factor increasing the computing time is that the time step of the imaginary-time-step method~\cite{Dav80,Reinhard82} implemented in the codes~\cite{Ev8article} has to be decreased with decreasing mesh size. This considerably slows down the convergence.
In Fig.~\ref{fig:Convergence} we show the evolution of the error on the total energy relative to \texttt{Lenteur} during the iterations for the nucleus \nuc{40}{Ca} for different mesh discretizations $dx$. The most accurate result after 100 iterations is obtained with $dx=1.0$~fm.
Gaining an order of magnitude of accuracy after convergence requires to carry out roughly 100 more iterations for the step sizes represented on the figure.

\subsection{Treatment of the long-range Coulomb interaction}

The direct Coulomb energy requires a special treatment because of its long range. One of the spatial integrations in Eq.~\eqref{eq:Ecoul} can be eliminated through the calculation of
the Coulomb potential of the protons, which
satisfies the electrostatic Poisson equation
\begin{equation}
\label{eq:PoissonEq}
\Delta U (\vec{r})
= - 4 \pi e^2 \rho_p(\vec{r}) \, ,
\end{equation}
where $e^2$ is the square of the elementary charge.
When solving this equation, boundary conditions need to be imposed at the edge of the box.These can be easily constructed when recalling that at large distances the potential is entirely determined by the multipoles of the nucleus' charge distribution $\langle \hat{Q}_{\ell m} \rangle$. Expanding the Coulomb potential on spherical harmonics and keeping terms up to \mbox{$\ell=2$}, the Coulomb potential outside the box is approximated by
\begin{equation}
\label{eq:BoundaryConditions}
U(\vec{r})
= \frac{e^2 Z}{r}
  + e^2 \frac{\langle \hat{Q}_{20} \rangle Y_{20}(\vec{r}) + \langle \hat{Q}_{22} \rangle \, \Re Y_{22}(\vec{r})}{r^3} \, ,
\end{equation}
which provides the boundary condition for the numerical solution of Eq.~\eqref{eq:PoissonEq}. The direct Coulomb energy is then calculated as
\begin{equation}
\label{Ecoul:2}
E_{\rm Coul}^d
= \frac{1}{2}\int \! d^3r \; U(\vec{r}) \, \rho_p(\vec{r}) \, .
\end{equation}
As for the nuclear part of the energy, the accuracy of the electrostatic potential, obtained by solving Eq.~\eqref{eq:PoissonEq}, is limited by three factors: the size of the box, the mesh discretization length $dx$ and the way derivatives are calculated.

A suitable box size for the Coulomb problem has to be larger than for the Skyrme EDF. This is a direct consequence of the long range of the Coulomb force. To make negligible the contributions to the boundary conditions of terms higher than $\ell=2$, see Eq.~\eqref{eq:BoundaryConditions}, one has to calculate the Coulomb potential in a box larger than the one used for the nuclear part of the interaction. Typical values are given in Table~\ref{tab:boxsize}. For light nuclei such as \nuc{40}{Ca}, no extra points for Coulomb need to be added, while the box has to be significantly enlarged for heavier systems in the \nuc{132}{Sn} and \nuc{208}{Pb} regions. For the calculation of the fission barrier of heavy nuclei such as \nuc{240}{Pu} up to very large deformations, the Coulomb box size has to be two times larger than the one needed for the Skyrme EDF to obtain the same nuclear accuracy on all the energies.

The Laplacian in Eq.~\eqref{eq:PoissonEq} has to be approximated on the mesh in such a way that the accuracy on the Coulomb energy is similar to the one of the other terms in the EDF. We show in Fig.~\ref{fig:coulcomparison}  the gain in accuracy on the total energy of \nuc{208}{Pb} obtained by going from a three-point to a seven-point FD formula for the Laplacian. Already a five-point formula brings the required accuracy and is used in all other calculations reported here. One can easily understand that a lower-order finite-difference formula than the one used to calculate the kinetic energy is sufficient for the Laplacian in Eq.~\eqref{eq:PoissonEq}: the typical length scale of the variation of the Coulomb potential is much larger than the scale on which the wave functions vary.

The final factor for the accuracy of the Coulomb solution is the mesh discretization length $dx$. As the effect of the Coulomb term is already incorporated in all of the applications, we will not discuss it separately.

\section{Discussion}

\subsection{Binding energies}
\label{sec:TotalE}

Provided that the box size is large enough, the main factor determining the accuracy of our implementation of mesh calculations is the discretization length.
In Fig.~\ref{fig:DxSphericals} the energy difference with respect to \texttt{Lenteur} results is plotted for three doubly-magic spherical nuclei, as a function of the mesh discretization $dx$ for a representative set of Skyrme parameterizations. It is remarkable that the interactions are grouped according to their effective mass (see Appendix~\ref{app:Forces} for the actual values): interactions with larger effective mass $m^*$ give systematically more accurate results than interactions with smaller ones. This property is related to the term $E_{\rho \tau}$ term of the Skyrme EDF in Eq.~\eqref{eq:skyrme:energy} that, in our experience, is the least well represented on a mesh. Since the magnitude of this term increases when the effective mass decreases, the accuracy obtained for a given mesh size deteriorates for lower effective mass.

One can see that the accuracy obtained with a mesh discretization as large as $dx=1.0$~fm is lower than $1.0$~MeV for \nuc{208}{Pb}. The energy difference decreases to a few hundred keV for $dx=0.8$~fm and to a few keV for $dx=0.6$~fm. Note that a similar accuracy for $dx=0.6$~fm was found  for a 2-dimensional code based on splines~\cite{PSF08a}. To obtain an agreement between the spherical code \texttt{Lenteur} and our 3-dimensional codes below the $1$~keV level would require to increase the box size but also to make the codes more similar. For a nucleus with a binding energy larger than 1 GeV, this implies a relative discrepancy better than $10^{-7}$ and there are several sources of differences in the codes that can play a role, none of which is easy to control.

\subsection{Deformation energy curves}

\begin{figure}[t!]
\includegraphics[width=.45\textwidth]{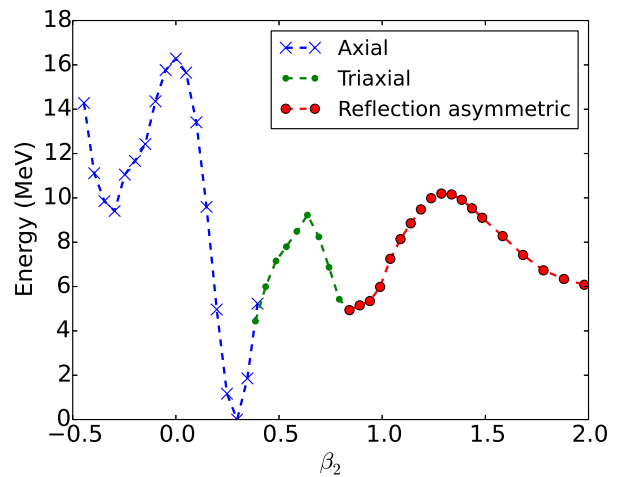}
\caption{(Color online) Energy curve for \nuc{240}{Pu} calculated with $dx=0.6$~fm. The regions where the deformation is axial, triaxial or axial and reflection asymmetric are indicated in blue, green and red respectively.}
\label{fig:PuLandscape}
\end{figure}

Let us now study the convergence properties of our numerical scheme for the fission path of \nuc{240}{Pu}. Our motivation is twofold: \nuc{240}{Pu} is a frequent benchmark for models that describe fission~\cite{Flocard74,Berger84,Blum94,Rutz95,Schunck14b,Younes09} but also for numerical algorithms \cite{PSF08a,Lu2014}. The energy curve of this nucleus presents two minima at prolate deformations, the ground state and a fission isomer. In Fig.~\ref{fig:PuLandscape}, we show the variation of the energy with deformation. 

\begin{figure}[t!]
\includegraphics[width=.45\textwidth]{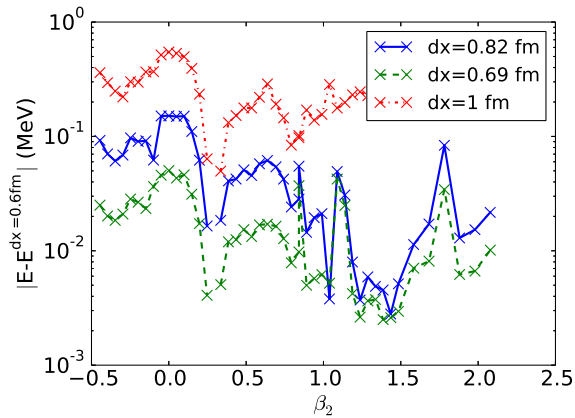}
\caption{(Color online) Energy differences between the results obtained for \nuc{240}{Pu} with $dx=1.0, 0.82$ and $0.69$~fm and those corresponding to $dx=0.6$~fm.}
\label{fig:Puerror}
\end{figure}

The box used for these calculations has the same size for all discretizations, as indicated by Table~\ref{tab:boxsize}. When the left-right symmetry is broken, the number of points along the $z$~direction is doubled. We have performed calculations with four different mesh discretization, $dx=1.0$, 0.82, 0.69 and 0.60~fm and tested the convergence as a function of $dx$ by taking the difference with respect to the results obtained with $dx=0.6$~fm. For each value of $dx$, the energy at each deformation is the energy relative to the prolate ground state.

The energy curve obtained with $dx=0.6$~fm is shown in Fig.~\ref{fig:PuLandscape}. The  topography obtained for other values of $dx$ is the same. Shapes are triaxial in the vicinity of the first barrier, whereas everywhere else they remain axial. At deformations smaller than the one of the fission isomer the configurations are reflection symmetric, whereas at larger deformations they are increasingly asymmetric.

We will use this curve as a reference to determine the accuracy of the calculations carried out for other values of $dx$. For each $dx$, the ground state energy is taken as the zero of the energy. The results are shown Fig.~\ref{fig:Puerror}.
The properties of the minimum are summarized in Table \ref{tab:PuStates}. The error decreases roughly by an order of magnitude by going from $dx=1.0$ to 0.82~fm and from $dx=0.82$ to 0.69~fm.
At $dx=1.0$~fm the error is of the order of a few times 100~keV, with a rather large oscillation. For a mesh discretization of $0.82$~fm, the error becomes lower than 100~keV (except in the vicinity of the spherical configuration where it reaches 150~keV, but this configuration is very excited) and is quite acceptable for the calculation of energy curves. Decreasing still the discretization to 0.69~fm reduces the error to values around a few times 10 keV at most.

Some published results allow for a comparison between the accuracies of mesh calculations and of calculations using an expansion on an HO basis. Pei \etal~\cite{PSF08a} have performed  calculations on an axial mesh using B-splines and on HO bases either spherical or deformed, with 20 oscillator shells in both cases.
The accuracy obtained in \cite{PSF08a} on a mesh of $dx=0.65$~fm seems very similar to the one we obtain. The use of a spherical HO~basis is rather unreliable, with an error larger than 1~MeV already for the excitation energy of the fission isomer and that quickly increases to several MeV at larger deformations. For an axial oscillator basis, the results are similar to those that we obtain with a mesh size of 0.82~fm up to the first barrier but the accuracy deteriorates rapidly for larger deformations, being of several hundreds of keV at the deformation corresponding to the fission isomer. Similar results can be found in \cite{NVR06a} for \nuc{194}{Hg} and in \cite{WER02a} for \nuc{256}{Fm}.

As a number of shells significantly larger than 20 is numerically prohibitive, one either has to resort to a two-center oscillator basis or one has to construct a suitable subspace within a much larger one-center HO basis by carefully selecting the low-lying single-particle states. The former option is developed in Ref.~\cite{Libert99} whereas the latter has been used during the construction of the
 unedf1 parametrization~\cite{KMN14a}, where the lowest 1771 basis states out of a basis of 50 HO shells has been kept. The accuracy obtained in this way for the excitation energy of the fission isomer is of the order of 100~keV. As a comparison, the experimental excitation energy of the fission isomer that can be found in the literature is 2.4 $\pm$ 0.3~MeV\cite{Bjorn80}, approximately 2.8~MeV \cite{Singh02} and 2.25 $\pm$ 0.20~MeV\cite{Hunyadi01}. In the light of these error bars, a numerical accuracy of 100~keV is sufficient for the adjustment of an EDF. However, from the published results by Pei \etal~\cite{PSF08a}, it can be estimated that the numerical error on the fission barrier height is a few times these $100$~keV. Similar results have been obtained in the case of the RMF method~\cite{Lu2014,ZLV15a}.

\begin{figure}[t!]
\includegraphics[width=.45\textwidth]{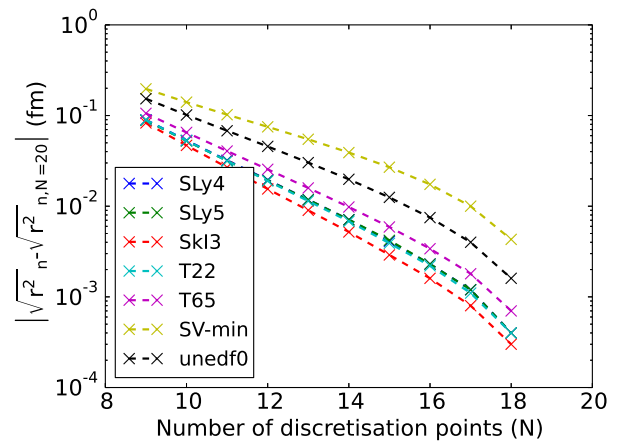}
 \caption{(Color online) Absolute difference in neutron rms radius for different Skyrme parameterizations for \nuc{34}{Ne} for $dx = 0.8$ fm. The reference calculation was performed in a box with $N=20$.}
 \label{fig:Ne34}
\end{figure}

\begin{figure}[t!]
\includegraphics[width=.45\textwidth]{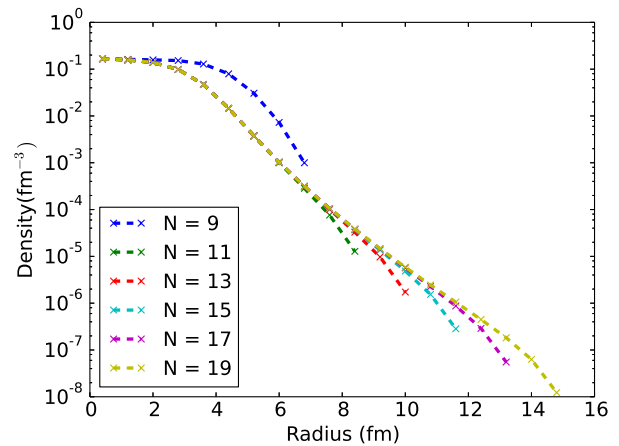}
 \caption{(Color online) Radial density profile of \nuc{34}{Ne} in different box sizes with $dx=0.8$~fm.}
 \label{fig:Ne34DensityProfile}
\end{figure}

\begin{figure*}
\includegraphics[width=.9\textwidth]{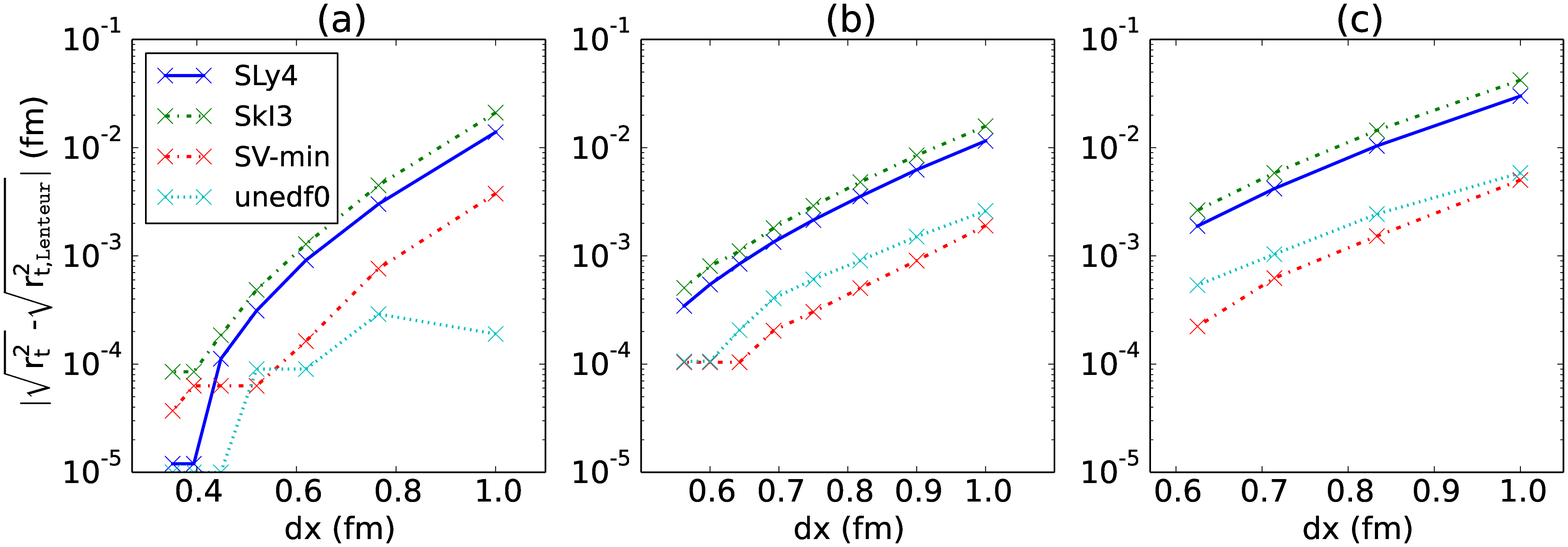}
\caption{(Color online) Absolute difference between the total rms radii calculated on a 3d mesh with respect to those of \texttt{Lenteur} as a function of the step size $dx$for the spherical nuclei \nuc{40}{Ca} (a), \nuc{132}{Sn} (b), and \nuc{208}{Pb} (c) and calculated with the Skyrme parameterizations as indicated.}
\label{fig:Radii}
\end{figure*}

\begin{figure}
 \includegraphics[width=.45\textwidth]{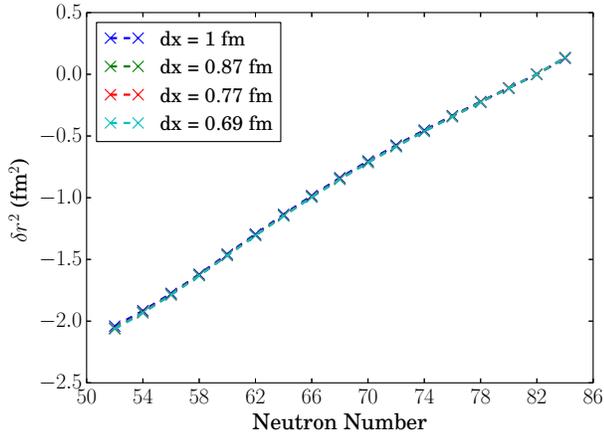}
 \caption{(Color online) Isotopic shifts $\delta r^2(N,Z)$ with respect to \nuc{132}{Sn} for different Sn isotopes and different mesh sizes.}
 \label{fig:SnShift}
\end{figure}

\subsection{Radial density distribution}
\label{sec:Radii}

The rms radius is intimately linked to the radial density distribution of a nucleus. One can expect that it is particularly sensitive to the box size for nuclei with a large excess of neutrons. Tests have been performed for the very neutron-rich nucleus \nuc{34}{Ne} by varying the box size for a fixed mesh discretization $dx=0.8$~fm. To avoid any ambiguity in the calculation, pairing has been omitted.
The results are presented in Fig.~\ref{fig:Ne34}, where we show the difference in total rms radius as a function of the box size for a representative set of EDF parameterizations.
For the size of the box recommended for \nuc{40}{Ca} in Table~\ref{tab:boxsize}, the number of points is 16 for a mesh size of 0.8~fm. It leads to an error of the order of $10^{-2}$~fm for most interactions, the results being slightly less accurate for SV-min. For smaller boxes, the accuracy of radii is lower and depends  on the interaction.

In Fig.~\ref{fig:Ne34DensityProfile} the radial profile of the total density of \nuc{34}{Ne} is plotted as a function of the box size. The distortion of the density in the smallest box is large and demonstrates that half the box size has to be larger than 8.0~fm. In all other boxes, the exponential tail of the density distribution is well described, up to the point before the last one. For a box size around 12~fm, the density is well described up to a decrease of the central density by six order of magnitudes.

The confinement in a volume is less evident in an expansion on a basis than in a mesh calculation, but it is also present. While oscillator basis functions extend to infinity, they are in practice strongly localized by their Gaussian form factor. If one takes its classical turning point as a measure for the extension of a HO state, one obtains, for \nuc{208}{Pb} and 20 oscillator shells, a value for the turning point that varies from 14~fm for $\ell=0$ to 16~fm for $\ell=20\hbar$. To increase the value of this turning point to 20~fm would require to use 28 oscillator shells for $\ell=0$.  This effect of confinement by an oscillator basis has been put in evidence in Ref.~\cite{Blazkiewicz2005} for the case for \nuc{112}{Zr}.

For comparison, the experimental uncertainty on rms charge radii for the Ne isotopes (up to $A=28$) varies from $0.002 \, \text{fm}$ close to stability to $0.02 \, \text{fm}$ for exotic isotopes~\cite{Angeli04}. It is interesting to note that the numerical accuracy of a mesh mean-field calculation has a similar level (provided the box is large enough), but that the model already introduces uncertainties on the rms radii that are at least one order of magnitude larger~\cite{Gao13}.

In Fig.~\ref{fig:Radii}, we compare the total rms radii calculated with decreasing mesh sizes to those obtained with \texttt{Lenteur} for three spherical nuclei \nuc{40}{Ca}, \nuc{132}{Sn} and \nuc{208}{Pb}. The agreement is already very satisfying for a large mesh size of 1.0~fm, with one order of magnitude gained in accuracy when decreasing the mesh size to 0.8~fm, which is the usual value of production calculations. An interesting feature that cannot be deduced from Fig.~\ref{fig:Radii} is that all of the parameterizations, with the exception of unedf0, always produce an rms radius that is smaller than the \texttt{Lenteur} result.

In Fig.~\ref{fig:SnShift}, we present the isotopic shifts $\delta r^2(N,Z)$ for a range of even-even Sn nuclei, the reference being \nuc{132}{Sn}. All curves almost exactly coincide. This demonstrates that the isotopic shifts are quite reliable even with coarse meshes. Similar results are obtained for Cd, Xe and Te isotopes.

\subsection{Two-neutron separation energies}

\begin{figure}
  \includegraphics[height=.22\textheight]{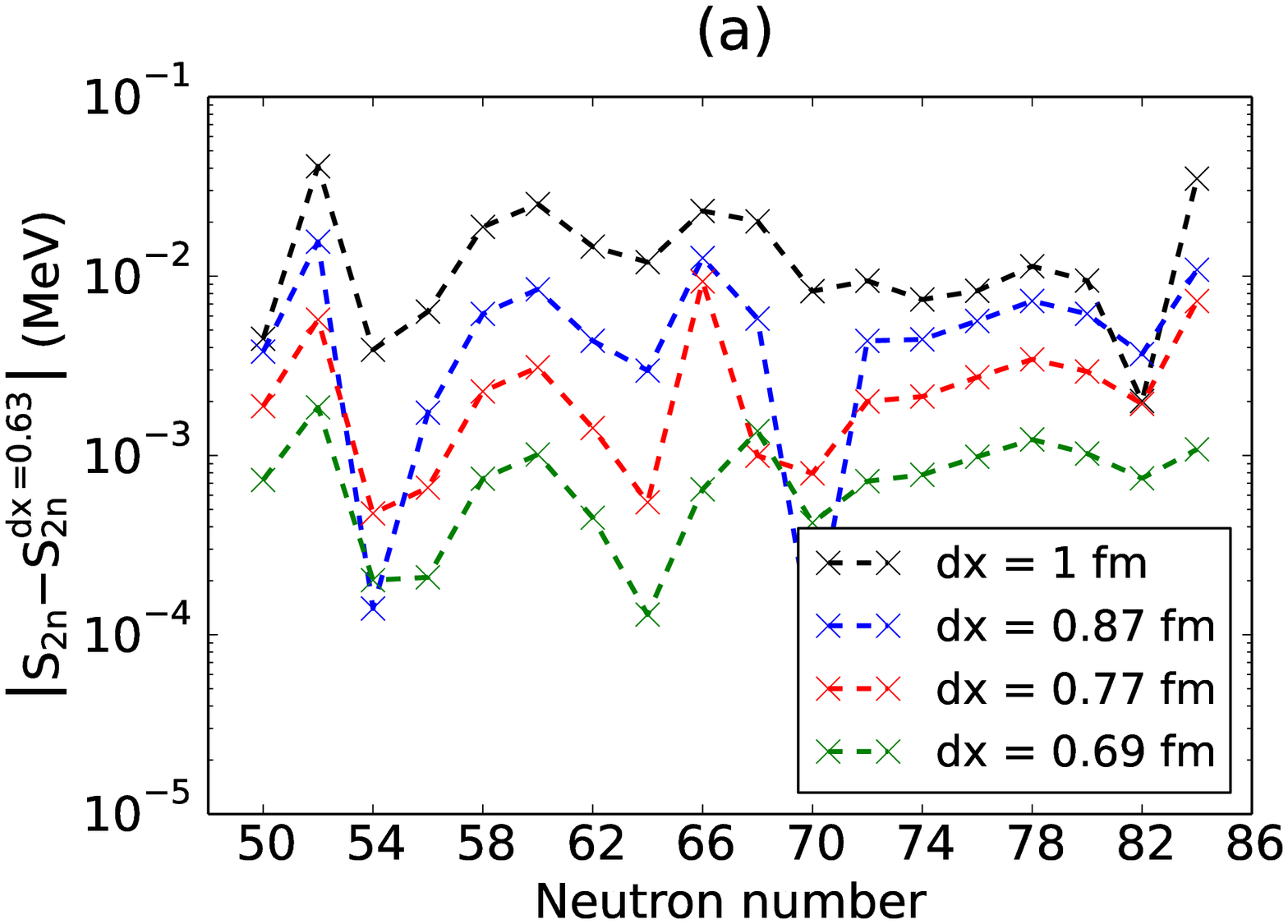}
  \includegraphics[height=.22\textheight]{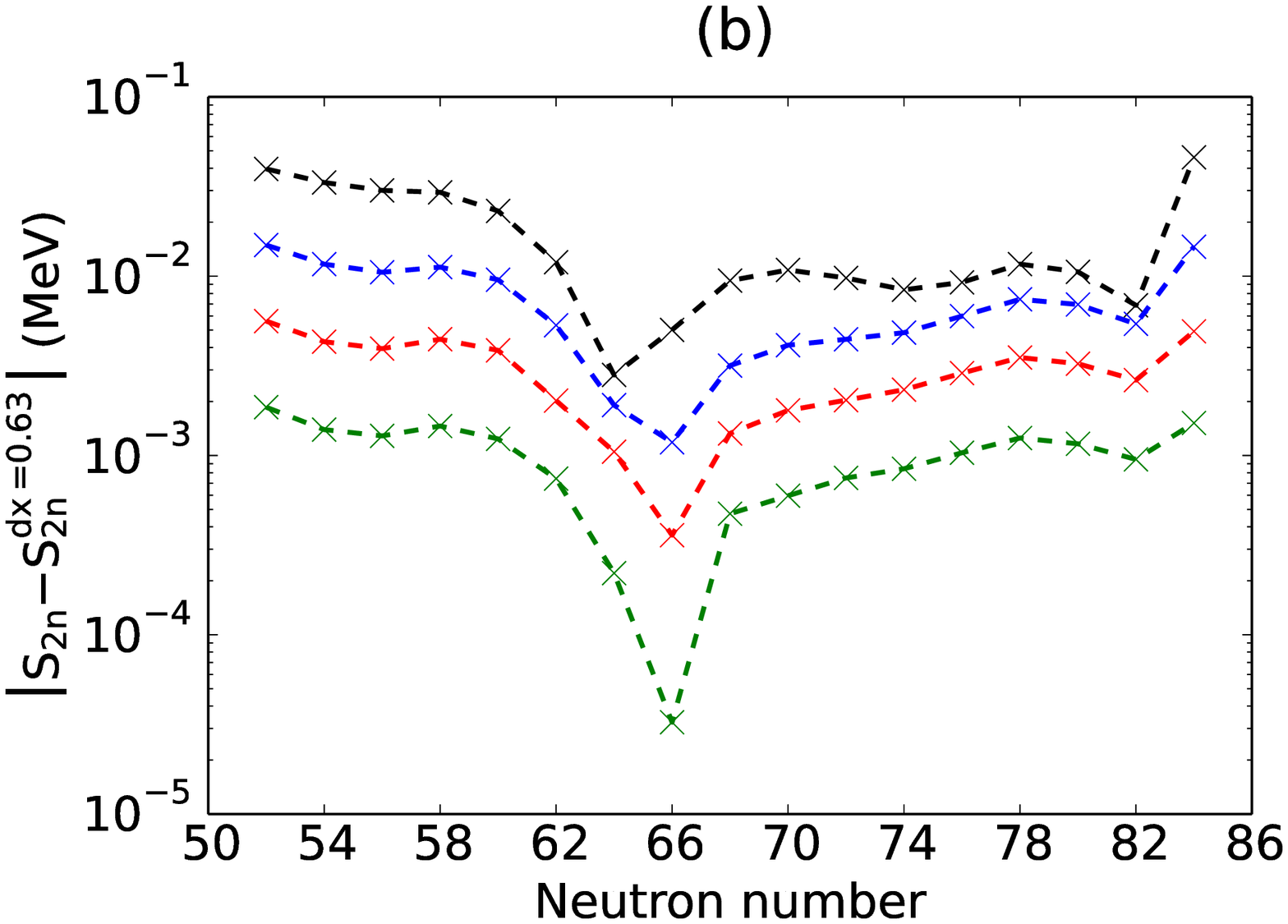}
  \includegraphics[height=.22\textheight]{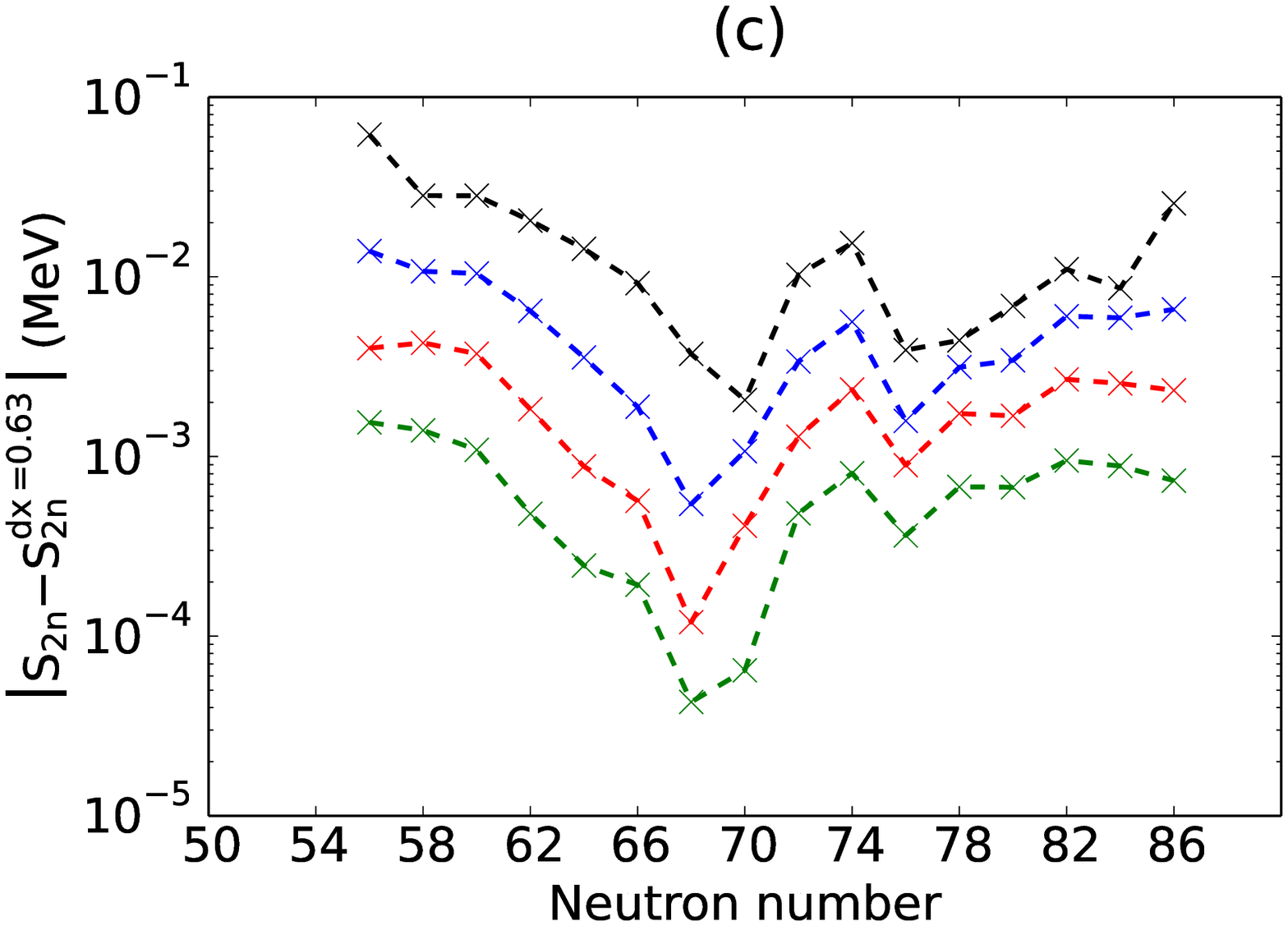}
  \includegraphics[height=.22\textheight]{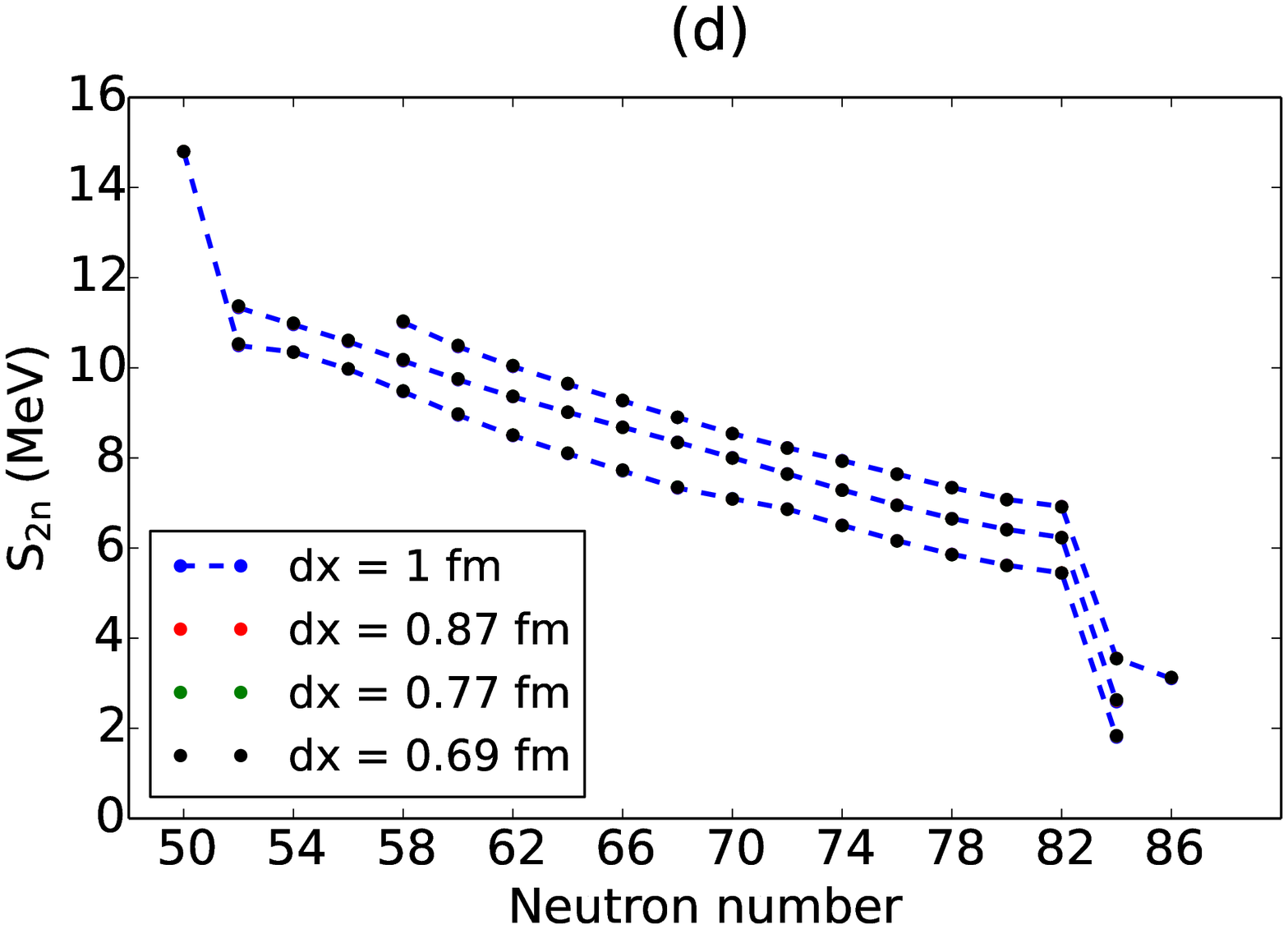}
 \caption{(Color online) Absolute differences of two-neutron separation energies between four mesh discretizations and a calculation with a mesh size of $dx=0.63$~fm for Cd (a), Sn (b), and Te (c) isotopes. In (d) we plot the two-neutron separation energies of all of these isotope chains as a reference.}
 \label{fig:S2Neutron}
\end{figure}

To put into evidence changes of nuclear structure with nucleon number, one often uses mass filters that are computed by taking specific differences between the binding energies of neighboring nuclei.
The simplest filter is the two-nucleon separation energy that is defined as the energy difference between two isotopes (or isotones) whose nucleon number differs by two.
In Fig.~\ref{fig:S2Neutron}, we show the evolution of the two-neutron separation energies, $S_{2n}$, of even-even nuclei for three neighboring isotopic chains when the mesh size $dx$ is decreased. For each discretization $dx$ we have plotted the difference of the $S_{2n}$ values to the one obtained at $dx=0.63$~fm.
Even with a mesh size as large as $dx=1.0$~fm, the accuracy on the $S_{2n}$ is already better than 100~keV, which is small enough for most applications. The mesh size used in most of our published applications, $dx=0.8$~fm leads to an accuracy better than 10~keV.
In the bottom panel, the two-neutron separation energies of the three isotope chains are plotted for four values of $dx$. The curves cannot be distinguished using a scale adapted to the variation of $S_{2n}$ as a function of the neutron number. This result is in strong contrast with respect to some published calculations using an expansion on an oscillator basis~\cite{Goriely11}, where special algorithms have to be devised to smooth numerical  irregularities that can be of the order of few hundred keV.

\begin{figure}
\includegraphics[width=.45\textwidth]{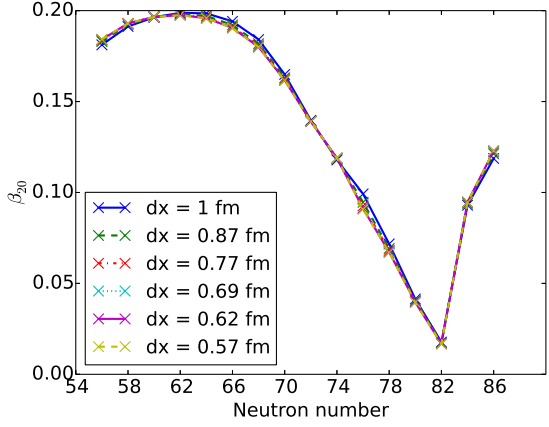}
\caption{(Color online) Mass $\beta_2$ quadrupole moment as a function of neutron number for a series of Te isotopes.}
\label{fig:TeBeta}
\end{figure}

\subsection{Multipole moments}
\label{sec:Multipole}

The dimensionless ground-state quadrupole moments $\beta_{2}$ of even-even Te isotopes are shown in Fig.~\ref{fig:TeBeta}. Differences between the curves corresponding to different values of $dx$ are tiny and not significant. Similar results were obtained for the Cd and Sn isotopes.

\begin{figure}
\includegraphics[width=.45\textwidth]{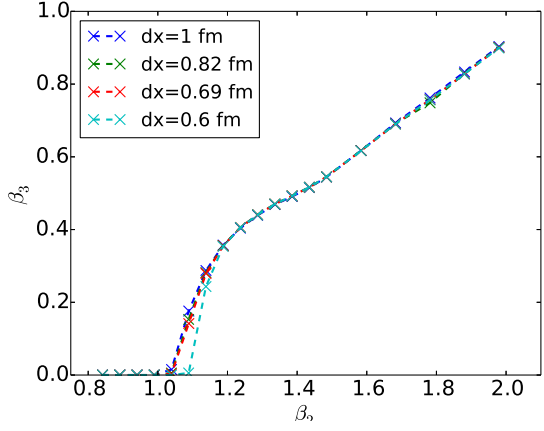}
\caption{(Color online) Mass octupole moment $\beta_{3}$ along the fission path for \nuc{240}{Pu}.}
\label{fig:Pu240Octu}
\end{figure}

\begin{figure}
\includegraphics[width=.45\textwidth]{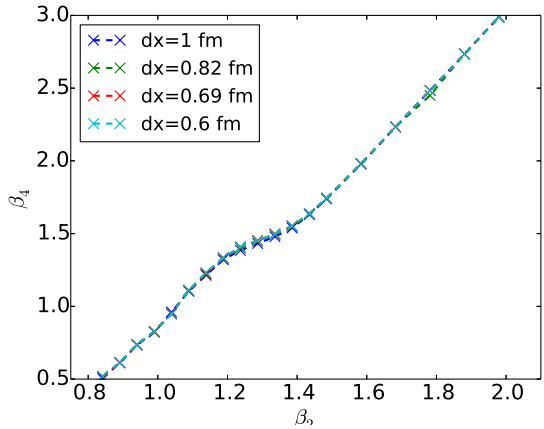}
\caption{(Color online) Mass hexadecapole moment $\beta_{4}$ for the parity breaking configurations along part of the fission path for \nuc{240}{Pu}.}
\label{fig:Pu240Hexa}
\end{figure}

We now examine how the multipole moments of \nuc{240}{Pu} along the fission path are affected by the mesh size. In Figs.~\ref{fig:Pu240Octu} and \ref{fig:Pu240Hexa} we show the octupole and hexadecapole moments, respectively, in the region of the fission path where parity is broken. Similar results obtained for the axial and triaxial cases are not shown.  In Tables~\ref{tab:PuStates} and~\ref{tab:PuIso} we show the multipole moments of the ground state and fission isomer of \nuc{240}{Pu} for the different mesh discretizations as obtained by unconstrained calculations.

From Figs.~\ref{fig:Pu240Octu} and~\ref{fig:Pu240Hexa} we see that the overall sequence of shapes along the fission path is robust with respect to the mesh spacing. The fission path is already precisely defined at the coarsest mesh ($dx=1.0$~fm) we used. A single exception can be seen at the onset of octupole deformation; in the vicinity of this point, however, the energy surface is very flat in $\beta_3$ direction.

On a smaller scale, the multipole moments do vary as a function of the mesh discretization. This is best visible in Tables~\ref{tab:PuStates} and~\ref{tab:PuIso}. Since our method hinges on the variation of the total energy in Eq.~\eqref{eq:Etot}, there is no guarantee that the values of the multipole moments converge in a predictable way. It is, however, reassuring to see that the typical variation of these moments is of the order of a few percent to at most about 10 percent. The larger variations present themselves in the higher-order $\beta_{6}, \beta_{8}$ and $\beta_{10}$ moments. These are more difficult to resolve on coarse meshes because of the high number of nodes their associated Legendre polynomials have.

 \begin{table}[t!]
 \begin{tabular}{ddcccdd}
  \hline\hline\noalign{\smallskip}
  \multicolumn{1}{c}{$dx$ (fm)} & \multicolumn{1}{c}{$E$ (MeV)} & \multicolumn{1}{c}{$\beta_2$} &  \multicolumn{1}{c}{$\beta_4$} & \multicolumn{1}{c}{$\beta_6$} & \multicolumn{1}{c}{$\beta_8$} & \multicolumn{1}{c}{$\beta_{10}$} \\
  \noalign{\smallskip}\hline\noalign{\smallskip}
     1.0  & -1801.909 & 0.288     & 0.160 & 0.043 & -0.002 &  -0.003 \\
   0.849  & -1802.770 & 0.289     & 0.163 & 0.045 &  0.001 &   0.001 \\
   0.739  & -1802.929 & 0.292     & 0.164 & 0.046 &  0.002 &   0.002 \\
   0.653  & -1802.969 & 0.290     & 0.165 & 0.046 &  0.002 &   0.001 \\
  \noalign{\smallskip}\hline\hline
  \end{tabular}
  \caption{Properties of the ground state of \nuc{240}{Pu} as obtained from
  non-constrained calculations.} 
  \label{tab:PuStates}
\end{table}

\begin{table}[t!]
 \begin{tabular}{ddccccc}
 \hline\hline\noalign{\smallskip}
  \multicolumn{1}{c}{$dx$ (fm)} & \multicolumn{1}{c}{$E$ (MeV)} & \multicolumn{1}{c}{$\beta_2$} &  \multicolumn{1}{c}{$\beta_4$} & \multicolumn{1}{c}{$\beta_6$} & \multicolumn{1}{c}{$\beta_8$} & \multicolumn{1}{c}{$\beta_{10}$} \\
 \noalign{\smallskip}\hline\noalign{\smallskip}
  1.0   & -1796.929 & 0.832 & 0.494  & 0.344 & 0.279 & 0.255 \\
  0.849 & -1797.950 & 0.840 & 0.510  & 0.367 & 0.303 & 0.278 \\
  0.739 & -1798.099 & 0.847 & 0.528  & 0.388 & 0.319 & 0.268 \\
  0.653 & -1798.123 & 0.841 & 0.516  & 0.375 & 0.312 & 0.259 \\
  \noalign{\smallskip}\hline\hline
  \end{tabular}
  \caption{Properties of the superdeformed fission isomer of as obtained from
  non-constrained calculations.} 
    \label{tab:PuIso}
\end{table}

\subsection{Single-particle levels}

\begin{figure}
 \includegraphics[width=.45\textwidth]{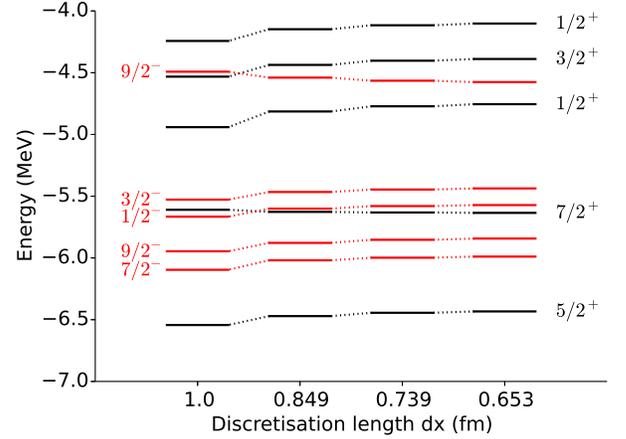}
 \caption{(Color online) Eigenvalues of the single-particle Hamiltonian for the ground state of \nuc{240}{Pu} as a function of mesh discretization $dx$. Only neutron single-particle levels within 1.5~MeV of the Fermi energy are shown.} 
\label{fig:PuLevels}
\end{figure}

In Fig.~\ref{fig:PuLevels} we show the evolution of the neutron single-particle levels within 1.5~MeV of the Fermi energy in the ground state of \nuc{240}{Pu} as a function of the mesh spacing $dx$. While slight shifts of the position of the levels are observed as a function of the mesh size, the largest error at $dx=1.0$~fm is of the order of 100~keV. One can also note that the level ordering within the parity subspaces is the same for all values of $dx$. A similar dependence on box parameters is found for the proton states and the lighter nuclei studied here.

\section{Conclusion}
 
The aim of this paper was to study the numerical accuracy of the solution of the self-consistent mean-field equations using a discretization on a 3-dimensional cartesian coordinate-space mesh.
 Three elements permit to control its numerical accuracy. The first one is the method used to calculate derivatives. Using Lagrange-mesh derivatives leads to much more accurate results than finite-difference formulas for derivatives. In addition, a cartesian Lagrange mesh corresponds to a representation in a closed subspace of the Hilbert space, such that it always provides an upper bound to the binding energy that becomes tighter when adding points outside a given box or when decreasing the distance of mesh points in a given box. Neither is the case for finite-difference derivatives.
However, we have shown quantitatively that the accuracy of a calculation that uses finite-difference formulas during the iterations can be significantly improved upon by recalculating the EDF at convergence with Lagrange-mesh derivatives. Again, this procedure provides us with an upper bound of the energy, thus restoring the variational character of the calculation. Using Lagrange derivatives during the iterations allows to still improve the accuracy on energies but at the cost of at least doubling the computing time.
 
The second element on which mesh calculations depend is the size of the box in which the nucleus is confined. The examples of doubly-magic nuclei and neutron-rich \nuc{34}{Ne} illustrate that results for energies and densities are already stable at small box sizes.

Thirdly, the quality of the results depends on the mesh size with errors on energies that are almost independent on the number of neutrons and protons and on the shape of the nucleus. A mesh size $dx=0.8$~fm guarantees an accuracy  that is in general better than 100~keV, which corresponds to a relative accuracy of less than a tenth of percent, even for lighter nuclei. Decreasing the mesh size to 0.7~fm permits to gain nearly an order of magnitude and to reach an accuracy that is well below all the uncertainties of the mean-field model.

One can summarize these results by concluding that a mesh technique as implemented in our codes is flexible (it can accommodate any kind of symmetry breaking), robust (the accuracy can be controlled by an adequate choice of the three elements mentioned above) and that it can be very accurate if needed. The positive aspect of our numerical scheme is that using a mesh size of 0.8~fm, as used in most of our past applications ensures an accuracy better than 100~keV on energies and reliable shape properties for nuclei of any mass.

Our study has been focussed on the solution of the mean-field equations and we have not touched the description of pairing correlations. There has already been a study of this problem by Terasaki \etal~\cite{Terasaki95}. It should be revisited today to take into account new developments. However, the problem is not exclusively a problem related to the way the mean-field equations are solved. The description of single-particle states well above the Fermi energy is probably very different when using a discretization on a mesh or an expansion on an oscillator basis.

\section*{Acknowledgements}
 The authors are grateful to many collaborators that have contributed to the codes used in this work, both with constructive criticism and sometimes their involvement with coding, in particular, P.~Bonche, H.~Flocard, J.~Meyer, J.~Dobaczewski, B.~Gall, N.~Tajima, J.~Terasaki, B.~Avez, B.~Bally and V.~Hellemans. Special thanks to K.~Bennaceur for providing us with \texttt{Lenteur}, B.~Bally for critical reading of the manuscript and G.~Scamps for pointing out a typographical error in the corrigendum to our Ref.~\cite{Ev8article}.

 This work has been supported in part by the European Union's Seventh Framework Programme ENSAR under grant agreement 262010 and by the Belgian Office for Scientific Policy under Grant No.\ PAI-P7-12.

\appendix


\section{Parameterizations used}
\label{app:Forces}

We here give some remarks on the interactions used throughout the text.
\begin{itemize}
 \item SLy4 and SLy5~\cite{Cha98}: these parameterizations were used as intended.
 \item T22 and T65~\cite{lesinski06a}: these parameterizations were used as intended.
 \item UNEDF0~\cite{unedf0}: while this parametrization was adjusted with a non-zero pairing interaction, we used it without any pairing.
 \item SV-min~\cite{klu09a}: this parametrization was adjusted with non-equal nucleon masses $m_n \neq m_p$. \texttt{Lenteur} does not handle this option, so we used instead the average value $\frac{m_n + m_p}{2}$ as nucleon mass.
 \item SkI3~\cite{reinhard95a}: this parameteriation was adjusted with the inclusion of a perturbative two-body c.m.\ correction, an option not included in \texttt{Lenteur}.
\end{itemize}
The values of their isoscalar effective mass are listed in Table~\ref{tab:effmass}.

\section{Pairing interaction}
\label{app:pairing}

The density-dependent pairing interaction we used for the isotope chains and the fission path of $^{240}$Pu is defined by~\cite{Rigollet99}:
\begin{equation}
\label{eq:v:ULB}
\hat{v}^{\text{pair}}(\vec{r},\vec{r}')
= - \frac{V_0}{2} \left( 1 - \hat{P}_{\sigma}  \right) \,
  \bigg[ 1 - \alpha \; \frac{\rho_0(\vec{R})}{\rho_s} \bigg] \,
  \delta (\vec{r}-\vec{r}') \, ,
\end{equation}
where $\rho_0 (\boldsymbol{R})$ is the isoscalar density at 
$\vec{R} = \tfrac{1}{2} ( \vec{r} + \vec{r}' )$. The parameters take
the values $\alpha = 1$, $\rho_s = 0.16 \, \text{fm}^{-3}$ and $V_0 = 1250 \, \text{MeV} \, \text{fm}^{-3}$.
In addition, this interaction was supplemented by two cutoffs, one above and the other below the Fermi energy, in order to eliminate the basis-size dependence of the total energy. They are defined by two Fermi functions
\begin{equation}
\label{eq:pair:cutoff:2}
f_k
= \big[ 1 + \text{e}^{(\epsilon_k - \lambda_q - \Delta \epsilon_q)/\mu_q}
  \big]^{-1/2}
  \big[ 1 + \text{e}^{(\epsilon_k - \lambda_q + \Delta \epsilon_q)/\mu_q}
  \big]^{-1/2} \, ,
\end{equation}
where $\lambda_q$ is the Fermi energy, $\epsilon_k$ the single-particle energy of the single-particle state $k$ and we chose $\mu_q = 0.5 \, \text{MeV}$, and $\Delta \epsilon_q = 5.0 \, \text{MeV}$ for protons and neutrons.

\section{The role of physical constant when using Skyrme EDFs}
\label{app:Constants}

By default, the physical constants used in our calculations are the following~\cite{Mohr10}:
\begin{eqnarray}
 e^2   & = & 1.43996446 \; \text{MeV fm}     ,                  \\
 m     =  \frac{m_n + m_p}{2} & = & 938.9187125 \; \text{MeV} \, c^{-2} , \\
 \hbar^2/(2m) & = & 20.735519104  \; \text{MeV fm}^2 \, .
\end{eqnarray}
Whenever possible, we used the value of $\hbar^2/(2m)$ that was used during the adjustment of the parametrization. It might seem superfluous to completely specify the physical constant used, but the results of our calculations depend on the precise values of these constants. In particular, the level of agreement between \texttt{Ev8} and \texttt{Lenteur} described in Sect.~\ref{sec:TotalE} is only attainable when these codes use exactly the same numerical values for the physical constants.

\begin{figure}[t!]
 \includegraphics[width=.4\textwidth]{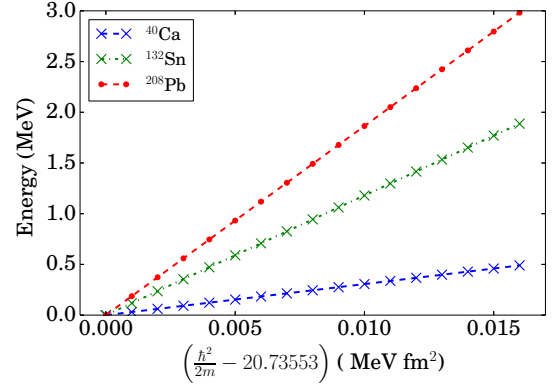}
 \caption{(Color online) Energy difference for the spherical nuclei \nuc{40}{Ca}, \nuc{132}{Sn} and \nuc{208}{Pb} for calculations with \texttt{Lenteur} using SLy4 with a modified value of $\hbar^2/(2m)$. The reference calculation is that obtained using the value used during the adjustment of SLy4.}
 \label{fig:hbm}
 \end{figure}

In fact, significant errors can be introduced when the values of the physical constants are slightly changed. The seemingly innocuous value of $\hbar^2/(2m)$ plays in fact a very important role. Figure~\ref{fig:hbm} shows \texttt{Lenteur} calculations for the spherical nuclei \nuc{40}{Ca}, \nuc{132}{Sn} and \nuc{208}{Pb} with SLy4. Every point was calculated by slightly changing the value of $\hbar^2/(2m)$ from $20.73553 \; \text{MeV fm}^2$, the SLy4 value. We see that using a value for $\hbar^2/(2m)$ that is not consistent with the value used during the fit of the EDF can lead to an error of several MeV on the total energy. $\hbar^2/(2m)$ is after all the proportionality constant of the kinetic energy in Eq.~\eqref{eq:Etot}. Typical values that have been used over the years vary at least between 20.73 and $20.7363 \; \text{MeV fm}^2$.
If the exact values of the physical constants used during the adjustment of a given parametrization are not available, then one cannot reliably compare the results with experimental data. In this case, one cannot judge the predictive power of this parametrization.

\begin{figure}[t!]
 \includegraphics[width=.4\textwidth]{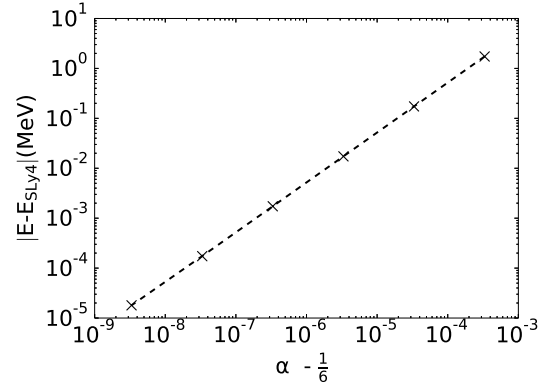}
 \caption{Difference between the total energy obtained with \texttt{Lenteur} 
for \nuc{40}{Ca} when using a rounded value for the density dependence parameter $\alpha$ in Eq.~\eqref{eq:skyrme:energy} instead of the full double precision value $\alpha = 1/6$ of the SLy4 parameteriation.}
 \label{fig:sigmadependence}
\end{figure}

Similar concerns arise for the parameters of the Skyrme interactions. The energy obtained in our calculations is more sensitive to some Skyrme parameters than to others, but the close agreement observed in Sect.~\ref{sec:TotalE} is not obtainable without carefully checking that the Skyrme parameters are completely consistent across codes. That this is not trivial can be concluded from Fig.~\ref{fig:sigmadependence}. There we plotted the relative difference in energy found by \texttt{Lenteur} between modified versions of the SLy4 functional and the correct SLy4. The interaction parameters are the same for every point, save for the density dependence parameter $\alpha$ in Eq.~\eqref{eq:skyrme:energy}. There are only very few parameterizations for which the value of $\alpha$ corresponds to a terminating decimal, for example SV-min for which $\alpha = 0.255368$. For the large majority of parameterizations the value of $\alpha$ is either $1/3$ or, as in the case of SLy4, $1/6$. Both of these correspond to a repeating decimal number, whose numerical representation might differ from code to code. Using $\alpha = 0.1667$ in a calculation with SLy4 corresponds to a rounding error of $\alpha - 1/6 \simeq 3.33 \times 10^{-5}$, which introduces an error of the total binding energy of \nuc{40}{Ca} of a few tens of keV.

It can clearly be seen that a limited representation of $\alpha$ implies a roundoff error that has a visible effect on the energy. This kind of error shows up when comparing \texttt{Lenteur} and \texttt{Ev8} results and for this reason we conclude that relative errors smaller than $10^{-5}$ become meaningless. Similar analyses can be made for the other interaction parameters, including the values of physical constants used to fit the interaction.

\begin{table}[t!]
 \begin{tabular}{lcd}
 \hline\hline\noalign{\smallskip}
  Parametrization & & \multicolumn{1}{c}{\text{$m^*/m$}} \\
  \noalign{\smallskip}\hline\noalign{\smallskip}
  SLy4 \& SLy5    & \cite{Cha98}       & 0.68  \\
  T22  \& T65     & \cite{lesinski06a} & 0.7   \\
  SkI3            & \cite{reinhard95a} & 0.577 \\
  SV-min          & \cite{klu09a}      & 0.95  \\
  unedf0          & \cite{unedf0}      & 0.9   \\
  \noalign{\smallskip}\hline\hline
 \end{tabular}
 \label{tab:effmass}
 \caption{Value of the isoscalar effective mass in units of the nucleon mass $m^*/m$ of the interactions used throughout this paper.}
\end{table}


\begin{thebibliography}{100}

	\bibitem{bender03a} M. Bender, P.-H. Heenen, and P.-G. Reinhard,
Self-consistent mean-field models for nuclear structure,
Rev. Mod. Phys. \textbf{75}, 121 (2003).

	\bibitem{Gao13} Y. Gao, J. Dobaczewski, M. Kortelainen, J. Toivanen and D. Tarpanov,
Propagation of uncertainties in the Skyrme energy-density-functional model,
Phys. Rev. C \textbf{87}, 034324 (2013).

	\bibitem{Dob14a} J. Dobaczewski, W. Nazarewicz, and P.-G. Reinhard,
Error estimates of theoretical models: a guide,
J. Phys. G \textbf{41}, 074001 (2014).

	\bibitem{JPGerrors} Focus Issue on
\textit{Enhancing the interaction between nuclear experiment
and theory through information and Statistics},
edited by D. G. Ireland and W. Nazarewicz,
J. Phys. G \textbf{42} (2015).

	\bibitem{BFH85a} P. Bonche, H. Flocard, P.-H. Heenen, S. J. Krieger, and M. S. Weiss,
Self-consistent study of triaxial deformations: application to the isotopes of Kr, Sr, Zr and Mo,
Nucl. Phys. \textbf{A443}, 39 (1985).

	\bibitem{Mar14a} J. A. Maruhn, P.-G. Reinhard, P. D. Stevenson, and A. S. Umar,
The TDHF code Sky3D,
Comp. Phys. Comm. \textbf{85}, 1410 (2014).

	\bibitem{Umar1991} A. S. Umar and M. R. Strayer,
Basis-spline collocation method for the lattice solution of boundary value problems,
Comp. Phys. Comm. \textbf{63}, 179 (1991).

	\bibitem{PSF08a} J. C. Pei, M. V. Stoitsov, G. I. Fann, W. Nazarewicz, N. Schunck, and F. R. Xu,
Deformed coordinate-space Hartree-Fock-Bogoliubov approach to weakly bound nuclei and large deformations,
Phys. Rev. C \textbf{78}, 064306 (2008).

	\bibitem{Pei2014} J. C. Pei, G. I. Fann, R. J. Harrison, W. Nazarewicz, Y. Shi, and S. Thornton,
Adaptive multi-resolution 3D Hartree-Fock-Bogoliubov solver for nuclear structure,
Phys. Rev. C \textbf{90}, 024317 (2014).

	\bibitem{Bay86a} D. Baye and P.-H. Heenen,
Generalised meshes for quantum mechanical problems,
J. Phys. \textbf{A19}, 2041 (1986).

	\bibitem{Imagawa03} H. Imagawa and Y. Hashimoto,
Accurate random-phase approximation calculation of low-lying states on a three-dimensional Cartesian mesh,
Phys. Rev. C \textbf{67}, 037302 (2003).

	\bibitem{Hashimoto13} Y. Hashimoto,
Time-dependent Hartree-Fock-Bogoliubov calculations using a Lagrange mesh with the Gogny interaction,
Phys. Rev. C \textbf{88}, 034307 (2013).

	\bibitem{Lu2014} B.-N. Lu, J. Zhao, E.-G. Zhao, and S.-G. Zhou,
Multidimensionally-constrained relativistic mean field models and potential energy surfaces of actinide nuclei
Rev. C \textbf{89}, 014323 (2014).

	\bibitem{HFBTHO1} M. V. Stoitsov, J. Dobaczewski, W. Nazarewicz, and P. Ring,
Axially deformed solution of the Skyrme-Hartree-Fock-Bogoliubov equations using the transformed harmonic oscillator basis (II) \texttt{HFBTHO} v2.00d: A new version of the program,
Comp. Phys. Comm. \textbf{167}, 43 (2005).

	\bibitem{Rodriguez2014} T. R. Rodr\'iguez, A. Arzhanov, and G. Mart\'inez-Pinedo,
Mean field and beyond-mean-field global calculations with Gogny interactions,
Phys. Rev. C \textbf{91}, 044315 (2015).

	\bibitem{Schunck14} N. Schunck, J. D. McDonnell, J. Sarich, S. M. Wild, and D. Higdom,
Error analysis in nuclear density functional theory,
J. Phys. G,  Nucl. Part. Phys. \textbf{42}, 034024 (2015).

	\bibitem{RER02a} R. Rodr\'iguez-Guzm\'an, J. L. Egido, and L. M. Robledo,
Correlations beyond the mean field in magnesium isotopes: angular momentum projection and configuration mixing,
Nucl. Phys. \textbf{A709}, 201 (2002).

	\bibitem{BFH05a} P. Bonche, H. Flocard, and P.-H. Heenen,
Solution of the Skyrme HF + BCS equation on a 3D mesh,
Comp. Phys. Comm. \textbf{171}, 49 (2005).

	\bibitem{Ev8article} W. Ryssens, V. Hellemans, M. Bender, and P.-H. Heenen,
Solution of the Skyrme-HF+BCS equation on a 3D mesh, II: A new version of the \texttt{Ev8} code,
Comp. Phys. Comm. \textbf{187}, 175 (2015);
Comp. Phys. Comm. \textbf{190}, 231(E) (2015).

	\bibitem{Olver} P. J. Olver,
Introduction to Partial Differential Equations,
(Springer, Cham, Heidelberg, New York, Dordrecht, London, 2014).

	\bibitem{Karim} K. Bennaceur, \textit{Lenteur HFB Code}, unpublished.

	\bibitem{Bul13a} A. Bulgac and M. McNeil Forbes,
Use of the discrete variable representation basis in nuclear physics,
Phys. Rev. C \textbf{87}, 051301(R) (2013).

	\bibitem{Blu92a} V. Blum, G. Lauritsch, J. A Maruhn, and P.-G. Reinhard,
Comparison of coordinate-space techniques in nuclear mean-field calculations,
J. Comp. Phys. \textbf{100}, 364 (1992).

	\bibitem{RutzThesis} K. Rutz,
\textit{Struktur von Atomkernen im Relativistic-Mean-Field-Modell},
Doctoral thesis, J. W. Goethe-Universit{\"a}t Frankfurt am Main
(Ibidem-Verlag, Stuttgart, 1999).

	\bibitem{Sza99a} V. Szalay,
Discrete variable representations of differential operators,
J. Chem. Phys. \textbf{99}, 1978 (1993).

	\bibitem{Sza12a} V. Szalay, T. Szidarovszky, G. Czak{\'o}, and A. G. Cs{\'a}sz{\'a}r,
A paradox of grid-based representation techniques: accurate
eigenvalues from inaccurate matrix elements,
J. Math. Chem. \textbf{50}, 636 (2012).

	\bibitem{Lit02b} R. G. Littlejohn, M. Cargo, T. Carrington, K. A. Mitchell, and B. Poirier,
A general framework for discrete variable representation basis sets,
J. Chem. Phys. \textbf{116}, 8691 (2002).

	\bibitem{Bay06a} D. Baye,
Lagrange-mesh method for quantum mechanical problems,
Phys. Stat. Sol. (b) \textbf{243}, 1095 (2006).

	\bibitem{Bay15a} D. Baye,
The Lagrange-mesh method,
Phys. Rep. \textbf{565}, 1 (2015).

	\bibitem{Fur12a} R. J. Furnstahl, G. Hagen, and T. Papenbrock,
Corrections to nuclear energies and radii in finite oscillator spaces,
Phys. Rev. C \textbf{86}, 031301(R) (2012).

	\bibitem{Coo12a} S. A. Coon, M. I. Avetian, M. K. G. Kruse, U. van Kolck, P. Maris,
and J. P. Vary,
Convergence properties of ab initio calculations of light nuclei
in a harmonic oscillator basis,
Phys. Rev. C \textbf{86}, 054002 (2012).

	\bibitem{Fur15a} R. J. Furnstahl, G. Hagen, T. Papenbrock, and K. A. Wendt,
Infrared extrapolations for atomic nuclei,
J. Phys. G \textbf{42}, 034032 (2015).

	\bibitem{Bay02a} D. Baye, M. Hesse, and M. Vincke,
The unexplained accuracy of the Lagrange-mesh method,
Phys. Rev. E \textbf{65}, 026701 (2002).

	\bibitem{Heenen93} P.-H. Heenen, P. Bonche, J. Dobaczewski, H. Flocard, S. J. Krieger, J. Meyer,
J. Skalski, N. Tajima, and M. S. Weiss,
Proc. of the International Workshop on 'Nuclear structure models',
Oak Ridge (U.S.A.), March 1992, World Scientific, Singapore, p. 3 (1993).

	\bibitem{HBD93a} P.-H. Heenen, P. Bonche, J. Dobaczewski, and H. Flocard,
Generator-coordinate method for triaxial quadrupole dynamics in Sr isotopes. (II) Results for particle-number projected states,
Nucl. Phys. \textbf{A561}, 367 (1993).

	\bibitem{Dav80} K. T. R. Davies, H. Flocard, S. Krieger, and M. S. Weiss,
Application of the imaginary time step method to the solution of the static Hartree-Fock problem,
Nucl. Phys. \textbf{A342}, 111 (1980).

	\bibitem{Reinhard82} P.-G. Reinhard and R. Y. Cusson,
A comparative study of Hartree-Fock Iteration Techniques,
Nucl. Phys \textbf{A378}, 418 (1982).

	\bibitem{Mohr10} P. J. Mohr, B. N. Taylor, and D. B. Newe,
CODATA recommended values of the fundamental physical constants: 2010,
Rev. Mod. Phys. \textbf{84}, 1527 (2012).

	\bibitem{Flocard74} H. Flocard, P. Quentin, D. Vautherin, M. V{\'e}n{\'e}roni, and A. K. Kerman,
Self-consistent calculation of the fission barrier of \nuc{240}{Pu},
Nucl. Phys. \textbf{A231}, 176 (1974).

	\bibitem{Berger84} J. F. Berger, M. Girod, and D. Gogny,
Microscopic analysis of collective dynamics in low energy fission,
Nucl. Phys. \textbf{A428}, 23 (1984).

	\bibitem{Blum94} V. Blum, J. A. Maruhn, P.-G. Reinhard, and W. Greiner,
The fission barrier of \nuc{240}{Pu} in the relativistic mean field theory
Phys. Lett. \textbf{B323}, 262 (1994).

	\bibitem{Rutz95} K. Rutz, J. A. Maruhn, P.-G. Reinhard, and W. Greiner,
Fission barriers and asymmetric ground states in the relativistic mean-field theory,
Nucl. Phys. \textbf{A590}, 690 (1995).

	\bibitem{Schunck14b} N. Schunck, D. Duke, H. Carr and, A. Knoll,
Description of induced nuclear fission with Skyrme energy functionals: Static potential energy
surfaces and fission fragment properties,
Phys. Rev. C \textbf{90}, 054305 (2014).

	\bibitem{Younes09} W. Younes and D. Gogny,
Microscopic calculation of \nuc{240}{Pu} scission with a finite-range effective force,
Phys. Rev. C \textbf{80}, 054313 (2009).

	\bibitem{Rigollet99} C. Rigollet, P. Bonche, H. Flocard, and P.-H. Heenen,
Microscopic study of the properties of identical bands in the $A = 150$ mass region,
Phys. Rev. C \textbf{59}, 3120 (1999).

	\bibitem{NVR06a} T. Nik\u{i}s\'{c}, D. Vretenar, and P. Ring,
Beyond the relativistic mean-field approximation: Configuration mixing of angular-momentum-projected wave functions,
Phys. Rev. C \textbf{73}, 034308 (2006).

	\bibitem{WER02a} M. Warda, J. L. Egido, L. M. Robledo, and K. Pomorski,
Self-consistent calculations of fission barriers in the Fm region,
Phys. Rev. C \textbf{66}, 014310 (2002).

	\bibitem{Libert99} J. Libert, M. Girod, and J.-P. Delaroche,
Microscopic descriptions of superdeformed bands with the Gogny force: Configuration mixing calculations in the $A \sim 190$ mass region,
Phys. Rev. C \textbf{60}, 054301 (1999).

	\bibitem{KMN14a} M. Kortelainen, J. McDonnell, W. Nazarewicz, P.-G. Reinhard, J. Sarich, N. Schunck, M. V. Stoitsov, and S. M. Wild,
Nuclear energy density optimization: Large deformations,
Phys. Rev. C \textbf{85}, 024304 (2012).

	\bibitem{Bjorn80} S. Bj{\o}rnholm and J. E. Lynn,
The double-humped fission barrier,
Rev. Mod. Phys. \textbf{52}, 725 (1980).

	\bibitem{Singh02} B. Singh, R. Zywina, and R. Firestone,
Table of superdeformed nuclear bands and fission isomers (Third edition),
Nucl. Data Sheets \textbf{97}, 241 (2002).

	\bibitem{Hunyadi01} M. Hunyadi, D. Gassmann, A. Krasznahorkay, D. Habs, P. G. Thirolf, M. Csatl\'os, Y. Eisermann, T. Faestermann, G. Graw, J. Guly\'as, R. Hertenberger, H. J. Maier, Z. M\'at\'e, A. Metz and, M. J. Chromik,
Excited superdeformed $K=0^+$ rotational bands in $\beta$-vibrational fission resonances of \nuc{240}{Pu},
Phys. Lett. \textbf{B505}, 27 (2001).

	\bibitem{ZLV15a} J. Zhao, B.-N. Lu, D. Vretenar, E.-G. Zhao, and S.-G. Zhou,
Multidimensionally constrained relativistic mean-field study of triple-humped barriers in actinides,
Phys. Rev. C \textbf{91}, 014321 (2015).

	\bibitem{Blazkiewicz2005} A. Blazkiewicz, V. E. Oberacker, A. S. Umar, and M. Stoitsov,
Coordinate space Hartree-Fock-Bogoliubov calculations for the zirconium isotope chain up to the two-neutron drip line,
Phys. Rev C \textbf{71}, 054321 (2005).

	\bibitem{Angeli04} I. Angeli,
A consistent set of nuclear rms charge radii:
properties of the radius surface $R(N,Z)$,
Atomic Data and Nuclear Data Tables \textbf{87}, 185 (2004).

	\bibitem{Goriely11} S. Goriely, N. Chamel, and J. M. Pearson,
HFB Mass Models for Nucleosynthesis Applications,
J. Kor. Phys. Soc. \textbf{59}, 2100 (2011).

	\bibitem{Terasaki95} J. Terasaki, P.-H. Heenen, H. Flocard, and P. Bonche,
3D solution of Hartree-Fock-Bogoliubov equations for drip-line nuclei,
Nucl. Phys. \textbf{A600}, 371 (1996).

	\bibitem{Cha98} E. Chabanat, P. Bonche, P. Haensel, J. Meyer, and R. Schaeffer,
A Skyrme parametrization from subnuclear to neutron star densities. Part II. Nuclei far from stabilities
Nucl. Phys. \textbf{A635}, 231 (1998); 
Nucl. Phys. \textbf{A643}, 441(E) (1998)

	\bibitem{lesinski06a} T. Lesinski, M. Bender, K. Bennaceur, T. Duguet, and J. Meyer,
Tensor part of the Skyrme energy density functional: Spherical nuclei,
Phys. Rev. C \textbf{76}, 014312 (2007).

	\bibitem{reinhard95a} P.-G. Reinhard and H. Flocard,
Nuclear effective forces and isotope shifts,
Nucl. Phys. \textbf{A584}, 467 (1995).

	\bibitem{klu09a} P. Kl{\"u}pfel, P.-G. Reinhard, T. J. B{\"u}rvenich, and J. A. Maruhn,
Variations on a theme by Skyrme: A systematic study of adjustments of model parameters,
Phys. Rev. C \textbf{79}, 034310 (2009).

	\bibitem{unedf0} M. Kortelainen, T. Lesinski, J. More, W. Nazarewicz, J. Sarich, N. Schunck, M. V. Stoitsov and S. Wild,
Nuclear energy density optimization,
Phys. Rev C \textbf{82}, 024313 (2010).


\end{thebibliography}
\end{document}